\documentclass[aps,twocolumn,superscriptaddress,floatfix,amsmath,a4paper]{revtex4-1}
\usepackage{graphicx}
\usepackage{ulem}

\usepackage{amsmath,amssymb}
\usepackage[utf8]{inputenc}
\usepackage{psfrag}

\newcommand{\rhoc}{\rho_{\scriptscriptstyle\rm c}}
\newcommand{\rhoa}{\rho_{\scriptscriptstyle\rm a}}
\newcommand{\rhod}{\rho_{\scriptscriptstyle\rm d}}
\newcommand{\rhoper}{\rho_{\scriptscriptstyle\rm per}}
\newcommand{\rc}{r_{\scriptscriptstyle\rm c}}

\begin{document}
\title{Spatial social dilemmas: dilution, mobility and grouping effects
with imitation dynamics}
\author{Mendeli H. Vainstein}
\affiliation{Instituto de F\'\i sica, 
Universidade Federal do 
Rio Grande do Sul, CP 15051, 91501-970 Porto Alegre RS, Brazil}
\author{Jeferson J. Arenzon}
\affiliation{Instituto de F\'\i sica, 
Universidade Federal do 
Rio Grande do Sul, CP 15051, 91501-970 Porto Alegre RS, Brazil}

\date{\today}

\begin{abstract}
We present an extensive, systematic study of the Prisoner's Dilemma and
Snowdrift games on a square lattice under a synchronous, noiseless imitation dynamics. 
We show that for both the occupancy of the network and the (random) mobility 
of the agents there are intermediate values that may increase the amount
of cooperators in the system and new phases appear. We analytically determine the 
transition lines between these phases 
and compare with the mean field prediction and the observed behavior on
a square lattice. We point out which are the more relevant 
microscopic processes that entitle cooperators to invade a population of
defectors in the presence of mobility and discuss the universality of
these results. 
\end{abstract}

\maketitle

\section{Introduction}
\label{section.introduction}

Spatially distributed viscous populations sustain 
cooperation due to the fact that individuals form clusters for self-defense and mutual 
support (see Refs.~\cite{DoHa05,Nowak06,SzFa07,RoCuSa09a,PeSz10} and references 
therein for reviews). Nonetheless, the conditions for the appearance and the 
properties of such cooperative regions are not fully understood both 
in real and model systems. Given that the spatial 
localization allows a continuing interaction within the local neighborhood, the population viscosity  
may prevent defectors from invading the whole population, what 
otherwise occurs under random mixing. Once the high viscosity 
constraint is relaxed and density permits~\cite{VaAr01,WaSzPe12}, agents are able 
to diffuse.  
There are many ways in which mobility~\cite{DuWi91,EnLe93,FeMi96,HaTa05,LeFeDi05} 
can be implemented: it may be random~\cite{VaSiAr07,JiZhYi07,SzSzSc08,DrSzSz09,SiFoVaAr09,YaWa11,SuKi11,GeCrFr13}, 
 strategy dependent~\cite{ChLiDaZhYa10}, driven by 
payoff~\cite{YaWuWa10,ChDaLiZhZhYa11,LiYaSh11,LiYaWuWa11},
success~\cite{HeYu09,Helbing09,Yu11}
or neighborhood~\cite{Aktipis04,JiWaLaWa10,Aktipis11,ChGaCaXu11a,ZhWaDuCa11,ZhZhWePeXiWa12}, 
take or not~\cite{MeBuFoFrGoLaMo09,ZhWaDuCa11,LiYaWuWa11} excluded volume into account, 
be local or long ranged, occur on a discrete lattice 
(regular or complex)~\cite{Koella00,VaSiAr07,JiZhYi07,SzSzSc08,SiFoVaAr09,DrSzSz09,JiWaLaWa10,YaWuWa10,Yu11,YaWa11}, 
in continuous space~\cite{MeBuFoFrGoLaMo09,ZhWaDuCa11,LiYaWuWa11}
or in a fully connected system, 
it may be explicit or included as a cost~\cite{Koella00}, 
etc. 
Our previous results~\cite{VaSiAr07,SiFoVaAr09} show that even in 
the simplest framework of random, non-contingent mobility of unconditional agents,
diffusion is remarkably able to enhance cooperation within broad
conditions. Besides the typical interval between generations, a new timescale
is involved when mobility is taken into account in this simple model, the
diffusion characteristic time. If the typical time a step takes to occur is much larger than the generation
interval, the high viscosity limit may be a reasonable approximation. On the other hand, if
diffusion is fast, the behavior should approach, density permitting, the fully mixed
case. An interesting regime is when both timescales are similar: in that case the order in
which the dynamics is performed,  whether
the offspring generation occurs before of after the diffusion step,
has important consequences for the cooperative outcome~\cite{VaSiAr07,SiFoVaAr09} and is
often neglected.

We consider a $2\times 2$ game with pure, unconditional strategies: 
cooperation (C) or defection (D). Cooperation involves a benefit to the
recipient at the expense of the provider. Depending on the mutual choice, 
the earned payoff is:  a reward $R$ 
(punishment $P$) if both cooperate (defect), $S$ (sucker's payoff) and 
$T$ (temptation) if one cooperates and the other defects, respectively.
In the Prisoner's Dilemma (PD) game, the above payoffs
are ranked as $T>R>P>S$ and $2R>T+S$. Thus, it 
clearly pays more to defect whatever the opponent's 
strategy: the gain will be $T>R$ if the other cooperates and $P>S$ in the
case of defection.
The dilemma appears since if both play D they get $P$, what is worse
than the reward $R$ they would have obtained had they both played C.
On the other hand, there are situations when mutual defection is even worst than
being exploited, and $P<S$. This defines a different game, in which $T>R>S>P$, 
known as Chicken or Snowdrift (SD)~\cite{Rapoport66}. 
Without loss of generality, we renormalize all values such that $R=1$ and $P=0$,
the values of $T$ and $S$ remaining as the parameters that define the
nature of the game.

In a randomly mating population (mean field limit) with both
C and D strategies present, defection will be the most rewarding strategy for
the PD game, 
independently of the opponent's choice. As shown by Nowak and May~\cite{NoMa92}, 
when spatial correlations are included in the population, for example by 
placing the agents on a lattice, cooperators form clusters in which the 
benefits of mutual cooperation can outweigh losses against defectors, 
thus enabling cooperation to be sustained, in contrast to the spatially 
unstructured game, where defection is favored (these effects of the spatial
structure may be due to either the distribution of agents in space or to the
context preservation during the dynamics, see \cite{RoCuSa09b} for a
detailed account). 
Since then, the original Nowak-May version was extended and modified in several 
different ways 
(see Ref.~\cite{SzFa07} and references therein).
Once placed on a network, by analyzing the possible neighborhoods~\cite{NoMa93,ScBeMu02}, 
one can divide the parameter space into several regions with different levels of
cooperation and spatial structures.
 In this work we extend their analysis
to include the SD game, dilution and mobility of the agents, locating all 
transitions between distinct phases. In the presence of defects (density
$\rho<1$) but 
without mobility, all transitions already present in the full system 
remain, but a few others appear because of the larger number of possible local 
configurations. Remarkably, when diffusive processes are also present, in which
an agent jumps to an empty site with probability $m$, whether 
new phases appear or not depends on the chosen dynamics.

A systematic study of how often spatial structure favors cooperative
behavior has been the program of a few papers (see, for example,
Refs.~\cite{Hauert02,ScBeMu02,SaPaLe06,ToLuGi06,RoCuSa09b,LiKeLiHu12}). The task is not simple due to the
multitude of different dynamical rules and lattice geometries that may
be considered~\cite{SzFa07}. Nonetheless, we complement these previous works by including 
dilution and mobility while considering a parallel imitation rule, in which 
each individual combats with all its closest neighbors (if any), accumulates
the corresponding payoff and then may either move or try to generate its
offspring. In the reproduction step,
each player compares its total payoff with those of its
neighbors and changes strategy, following the one with the
greatest payoff among them. This strategy changing updating rule preserves 
the total number of individuals, thus keeping $\rho$ constant. Notice that
although there is no noise in this updating rule, the random mobility to
be considered here now has a similar role and prevents the system from
becoming stuck on shallow minima.
Initially, an equal number of cooperators and
defectors are randomly placed on a two dimensional square lattice of linear
size $L$ and periodic boundary conditions, and the system is allowed to
evolve until a stationary state is attained, when the measures are
thus taken. 

The paper is organized as follows. In the next section, the phase
diagrams for fully occupied and diluted (with and without random
mobility) are obtained and compared with numerical simulations and
the mean field prediction. We then discuss the possible mechanisms
leading to the enhancement or inhibition of cooperation and
finally present our conclusions.

\section{Phase diagrams}
\label{section.phase}

\begin{figure}[tbh]
\includegraphics[width=8cm]{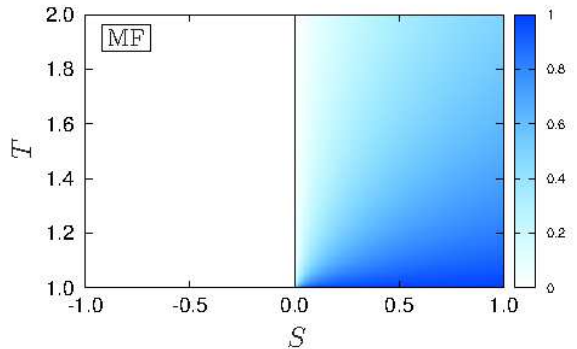}
\caption{Mean field (fully mixed) phase diagram 
for $P=0$ and $T>R=1$. The 
solid line shows the transition from the defector
dominated phase ($S<0$, PD game) to the coexistence one ($S>0$, SD
game). The scale at the right indicates the density
of cooperators, $\rhoc=S/(S+T-1)$, for $S>0$, with darker colors assigned to 
larger $\rhoc$.}
\label{phase_diagram_mf}
\end{figure}

When spatial correlations are not relevant, as when all agents interact
with all others (mean field limit), the phase diagram is easily obtained,
Fig.~\ref{phase_diagram_mf} (see, e.g., Ref.~\cite{Hauert02}).
For $P=0$ and $R=1$ there are two transition lines, one at $T=1$ and
other at $S=0$, dividing the $TS$ plane into two sections above 
$T=1$~\footnote{There are indeed two more regions for $T<1$, but they are
not considered here: for $S>0$ (Harmony game), $\rhoc=1$, and for $S<0$
(Stag Hunt game), the amount of cooperators depends on their initial
density: if it is larger (smaller) than $S/(S+T-1)$, then $\rhoc=0$ (1).},
the PD game for $S<0$ and the SD for $S>0$.
When $S<0$ (and $T>1$), defectors dominate and the relative~\footnote{In this
paper, all densities are relative to the total number  
of agents, not sites.}
density of cooperators, $\rhoc$,
is equal to 0. On the other hand, for 
$S>0$ (and again $T>1$), cooperators and defectors coexist with $\rhoc=S/(S+T-1)$. 

When spatial localization becomes an important factor, the corresponding
phase diagram can be constructed by analyzing all
the neighborhood configurations that are possible in the confrontation 
between two agents having different strategies. The transitions present in the phase diagram consider all
such configurations, irrespective of their probability of occurrence.  If a given
local configuration is rather rare, it might happen that our finite time simulations on a finite 
lattice are not able to sample it and, as 
a consequence, two phases might look rather similar or perhaps identical.
In the following, we consider a square 
lattice with the von Neumann neighborhood (nearest neighbors only) and 
without self-interaction, but the results can be extended to 
other lattices and neighborhoods, although the complexity of the task may vary. Unless specified,
all simulated systems have a linear length of $L=100$, and results are averaged over $100$ different initial random 
configurations  such that $\rhoc(0)=\rhod(0)=\rho/2$. 
The number of initial steps neglected before the asymptotic state depends on the density and mobility. 
In order to compare with our previous works, we consider the ``imitate-the-best'' dynamics, in which all individual's strategies 
are synchronously replaced by the strategy adopted by the individual with the highest collected payoff in the neighborhood. Besides this synchronous updating of the strategies, a Monte Carlo Step (MCS) also comprises an
attempt, by each agent, to diffuse: each agent blindly chooses a neighboring
site and, if it is empty, jumps to it with probability $m$. 
Using the notation of Schweitzer {\it et
al.}~\cite{ScBeMu02}, $K_{\theta}^n$ denotes the 
local occupation pattern and $p(K_{\theta}^n)$ the payoff acquired by an individual in such a configuration. 
Here, $n \in \{0,1,2,3,4\}$ gives the total 
number of cooperators in the local neighborhood and $\theta \in \{0,1\}$ 
describes whether the center cell is occupied by a defector or a cooperator, 
respectively. In the absence of empty sites, the number of defectors in
a neighborhood is $4-n$. The construction of the phase diagram amounts 
to the analysis of all the possible confrontations of $K_{0}^{n_0}$ and 
$K_{1}^{n_1}$, for $n_0 \in \{1,2,3\}$ and $n_1 \in \{0,1,2,3,4\}$. The 
value of $n_0=0$ is not important because a D having four D neighbors will 
either be surrounded by Ds with the same payoff or with higher payoff 
D players that have C neighbors, since $T>P$. The value of $n_0=4$ is 
not considered because with the chosen payoffs ($T>R$), a D with four C 
neighbors always has the highest possible payoff. This value might play a 
role when considering games in which $(T<R)$, as in the Stag 
Hunt~\cite{Hauert01,Hauert02}.

\subsection{Full occupancy ($\rho=1$)}

\begin{figure}
\includegraphics[width=8cm]{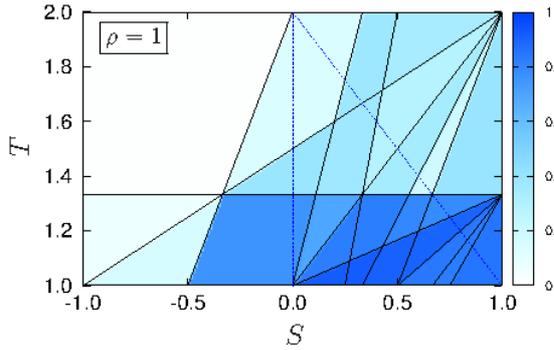}
\caption{Two dimensional cross section of the phase diagram displaying the asymptotic density of cooperators, for full occupancy
($\rho=1$) and $P=0$, $T>R=1$
and $S<R$. The solid lines represent the functions $f_{n_0n_1}$ which delimit
different phases (functions having the same $n_0$ intercept at the same point
on the line $S=1$). 
White regions are dominated by defectors, while the blue/grey ones have
some fraction of cooperators, its density indicated by the scale on the right.
The color code does not represent what happens at the transition lines, where draws happen and the fraction
of cooperators may differ from the two neighboring phases. 
Notice that, for a fixed $T$, $\rhoc$ is not monotonic in $S$ (see Fig.~\ref{fig.rho1}).
The dotted line $T=2-S$ is a usual choice for the payoff matrix: 
$T=1+r$ and $S=1-r$, $r$ being a payoff parameter. Another traditional
choice, also shown as a dotted line, is $S=0$ and separates the PD and SD games
(however, this is a transition line in the mean field case). The horizontal line
at $T=4/3$ separates the low C region (above) from the high C one (below).
}
\label{fig.phase_diagram}
\end{figure}

\begin{figure}
\includegraphics[width=8cm]{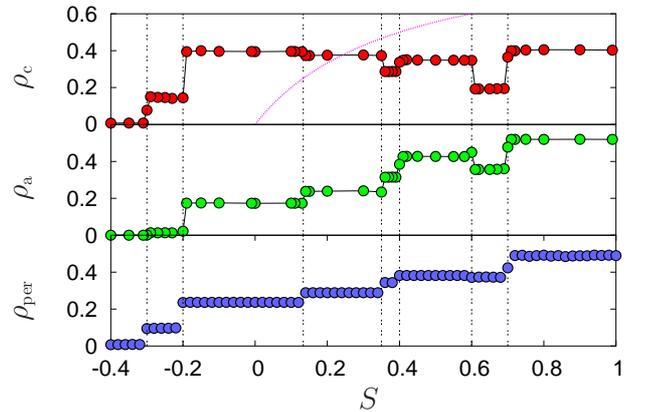}
\caption{(Top) Asymptotic density of cooperators as a function of $S$ for $T=1.4$, when the lattice 
is fully occupied ($\rho=1$) and the initial state is random. Notice both the abrupt change of $\rhoc$
when a transition line is crossed and the non monotonic behavior of $\rhoc$. On those lines, 
the value of $\rhoc$ may be very 
different from the neighboring phases. Also shown is the fully-mixed result, for
which the density of cooperators is non zero only for $S>0$ and is an increasing
function of $S$ in that interval. Notice that for the region
around the weak PD ($S=0$), spatial correlations enhance cooperation~\cite{NoMa92}.
(Middle) Fraction of active sites $\rhoa$ for the same parameters. 
Besides the transition points, only the $0.6<\rho<0.7$ region breaks the increasing monotonicity.
(Bottom) The fraction $\rhoper$ of cooperator-defector pairs
 (to be compared with the snapshots of Fig.~\ref{fig.rho1.snapshots}).}
\label{fig.rho1}
\end{figure}

We initially consider the simpler case without empty sites ($\rho=1$), in which
 obviously no mobility, as implemented here, is possible. For the PD  
and SD games, the following family of functions 
compare the payoffs that result from all possible confrontations 
of a D and a C, having local neighborhoods given by $K_0^{n_0}$ and 
$K_1^{n_1}$, respectively,
\begin{equation}
f_{n_0n_1}=\frac{n_1 R+(4-n_1)S-(4-n_0)P}{n_0}.
\label{eq.diag}
\end{equation}
 If $T>f_{n_0n_1}$, then the D with local configuration $K_{0}^{n_0}$
 will beat the C with local configuration  $K_{1}^{n_1}$; if
 $T<f_{n_0n_1}$, then the D will beaten; and if $T=f_{n_0n_1}$, there 
will be a draw between the two players. Indeed, the behavior at a 
transition point may be different from the neighboring regions, what 
makes each segment between line crossings a phase in itself. The phase 
diagram will then be composed of the regions defined by all these
 functions, together with the inequalities that define the games.
For given values of $R$, $P$ and $n_i$ in the allowed ranges, these 15 functions, 
which can be separated into 3 groups depending on $n_0$, represent the values of $T$ 
where there are transitions which divide the parameter space into different phases. Without 
loss of generality, we can take $R=1$, so that the above functions will describe planes 
in the three dimensional space of $T$, $P$ and $S$, leading to a complex phase diagram in three
dimensions. The usual choice of $P=0$ takes a 2d cross-section of this three dimensional
parameter space  allowing a simpler description of the phase diagram, 
as shown in Fig.~\ref{fig.phase_diagram}.
Within each phase, the behavior is the same, and it is enough to numerically
study a single representative point for synchronous imitation dynamics, as can be seen in 
Fig.~\ref{fig.rho1}, where different simulation points in the same phase lead to the same outcome. 
The transition between two neighboring phases is
usually discontinuous, the density of cooperators presenting abrupt jumps when a 
transition line is crossed, as can be seen in Fig.~\ref{fig.rho1}. 
\begin{figure*}[tbh] 
\includegraphics[width=3.5cm]{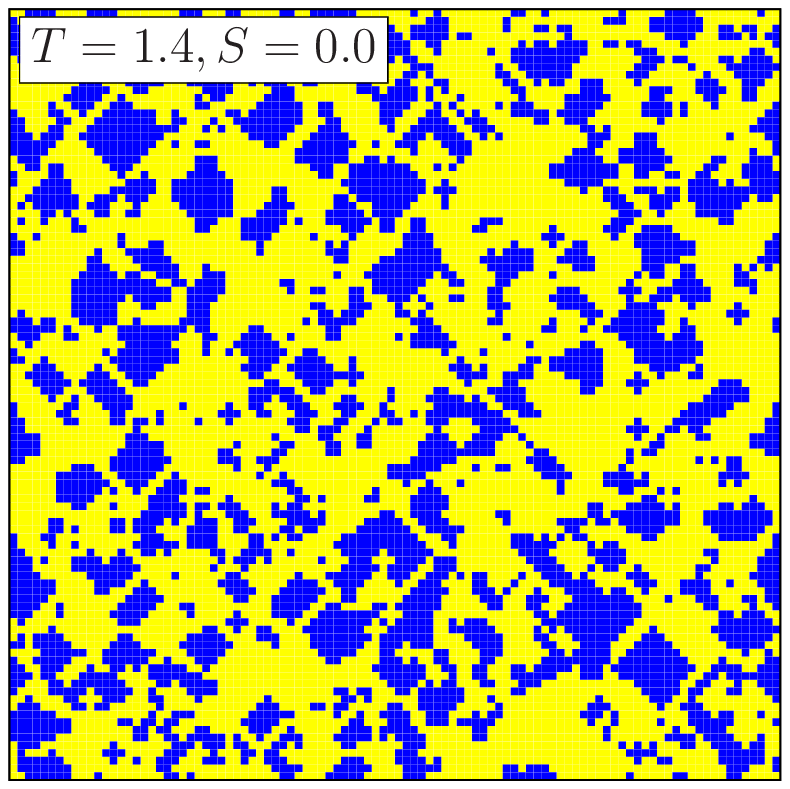}
\includegraphics[width=3.5cm]{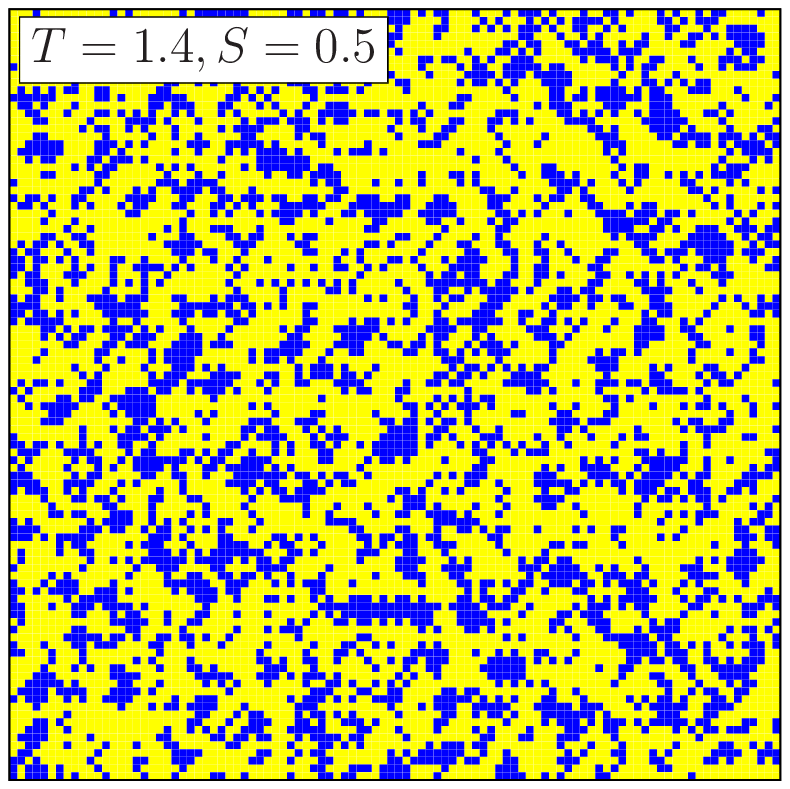}
\includegraphics[width=3.5cm]{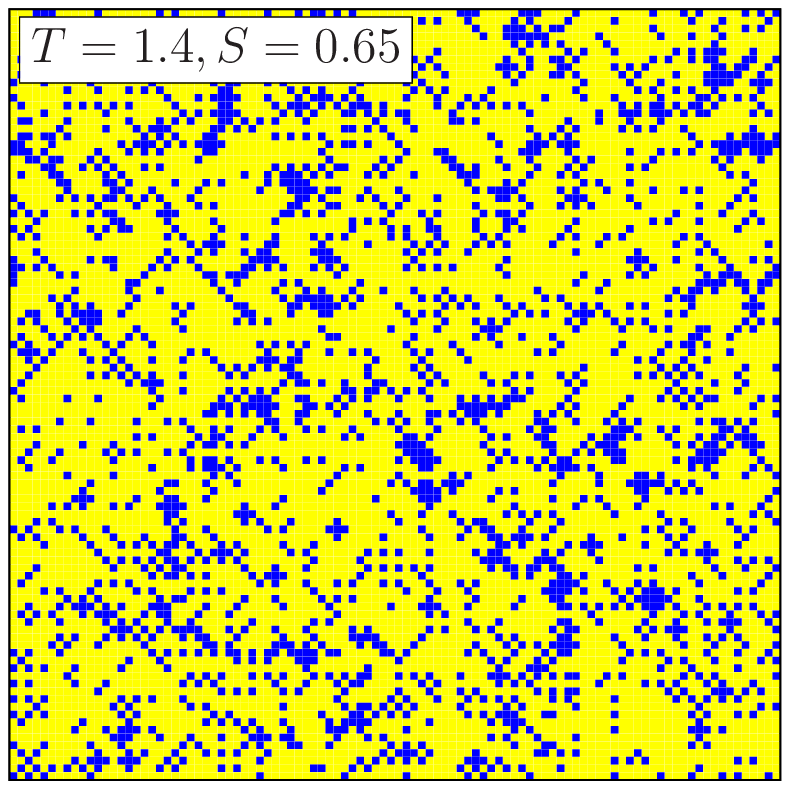}
\includegraphics[width=3.5cm]{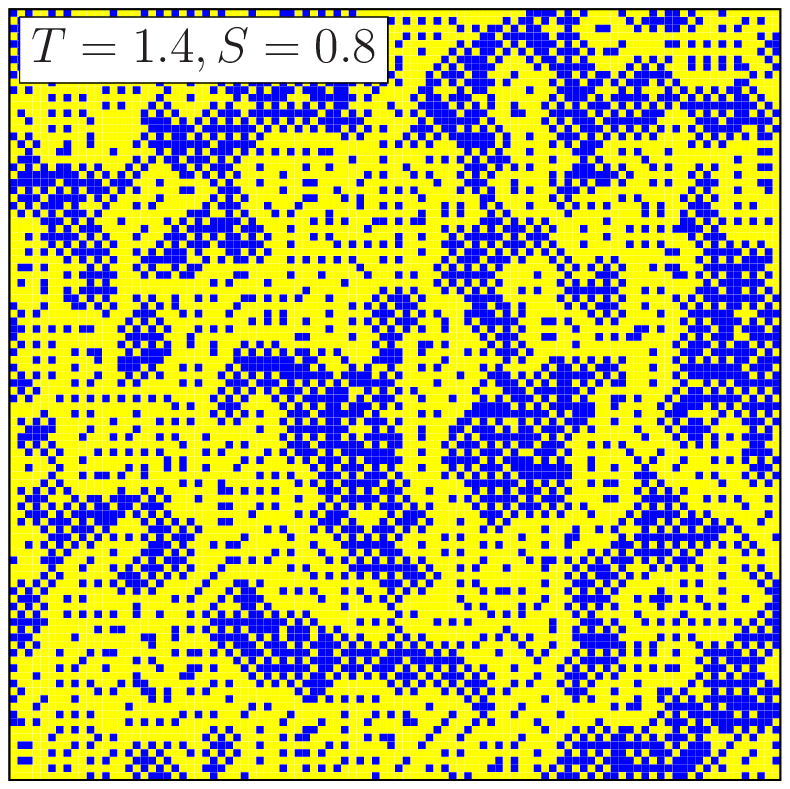}
\includegraphics[width=3.5cm]{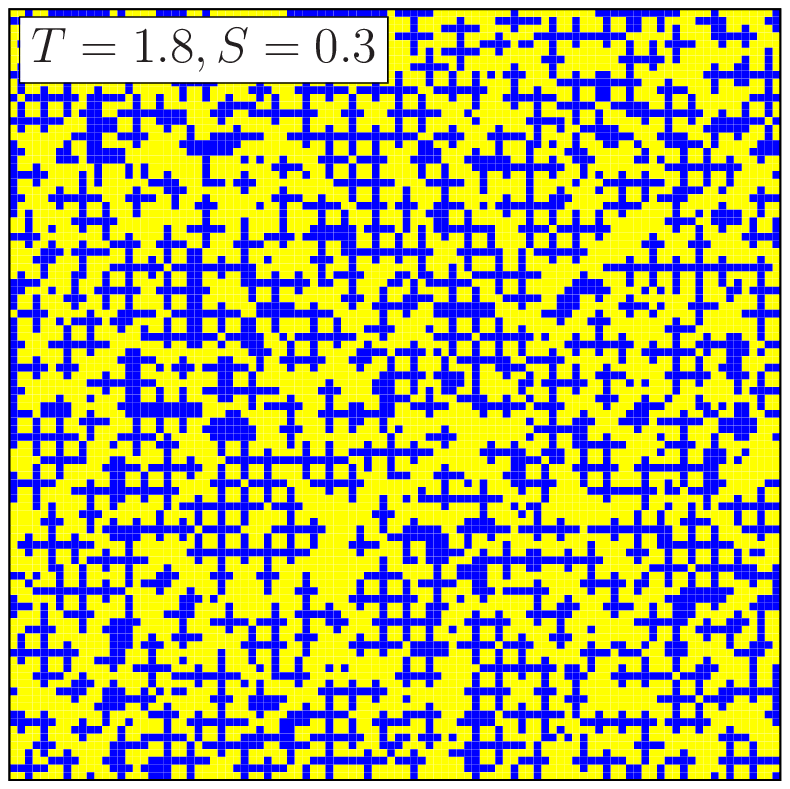}
\caption{Snapshots after $10^3$ MCS showing typical configurations for several values
of $T$ and $S$ for $\rho=1$. Blue/yellow (black/grey) sites represent cooperators/defectors,
respectively. Larger levels of cooperation are
associated with the presence of a long tail in the group size distribution, 
while the compactness of the large cooperator groups depends on $S$, being large (small) 
for small (large) $S$. Notice that the rightmost snapshot is the only one
with $T=1.8$. In this last case, $2\times 2$ squares of defectors separated by lines of
cooperators form a fully stable structure since $p(K_1^4)>p(K_0^2)>p(K_1^2)$. Because of the
random initial state, we only observe patches of such structure.}
\label{fig.rho1.snapshots}
\end{figure*}
 The dashed lines 
 in Fig.~\ref{fig.phase_diagram} are not transition lines, but two common parametrizations of the
payoff matrix for these games. The diagonal dashed line considers $T=1+r$ and $S=1-r$
(such that $T+S=2$), where $r$ is a parameter. The second parametrization line, the vertical one at $S=0$,
is exactly at the border between the SD and PD games, and is known as the weak PD.
Notice that several phases are left out with such parametrizations.

Fig.~\ref{fig.rho1} (top) shows, for $T=1.4$, the fraction of cooperators
as a function of $S$, along with the mean field result. Two important
features can be noticed. First of all, while for $S<0$ the spatial correlations 
significantly increase the amount of cooperation when compared to the
mean field limit, this is not always the case for $S>0$ (SD). Indeed, for $S\gtrsim 0.24$,
the mean field curve lies above the lattice results, while for $0<S\lesssim 0.24$,
spatial correlation improves cooperation. The second evident feature is the
non-monotonicity of $\rhoc$: as $S$ increases, one would intuitively expect
larger levels of cooperation; instead, some regions (most prominently, around
$S=0.65$) present a smaller than expected fraction of cooperators. 
A large amount of cooperation 
is related to the existence of a long tail in the distribution of group sizes
(a group is defined as a set of neighboring, same strategy agents), as is the
case, for example, for $S=0$ and $0.8$. On the other hand, for $S=0.65$ the system has 
a much lower level of cooperation and very few large clusters.
 This can be checked in Fig.~\ref{fig.rho1.snapshots} in which some characteristic
snapshots are shown for $T=1.4$ and 1.8.
For $S=0$, leftmost snapshot, the minimal cooperative cluster able to grow is 
$2\times 2$, while smaller or linear clusters are removed in the first steps of 
the dynamics. The surviving clusters are far away from each other and grow through
flat edges (with at least two cooperators) while diagonals are stable
(although both cooperators and defectors at a diagonal interface have two 
neighboring cooperators, and $p(K_1^2)<p(K_0^2)$, the cooperators are backed up by interior 
cooperators with higher payoffs). Thus, rather large and compact clusters 
may grow before starting to interfere with each other. Once they get close enough,
defectors trapped between these clusters will have cooperators at both sides, and therefore  will
acquire a large payoff and reproduce. These defector clusters will grow as well
until a dynamical equilibrium is achieved.
 For $S=0.5$, second snapshot, since $p(K_1^0)>p(K_0^1)$, single cooperators are able to seed a
growing cluster and survive in a sea of defectors. The clusters are much less compact than
in the previous case and the lattice is
populated by those smaller clusters that were decimated in the $S=0$ case. 
 Being less compact, clusters increase the amount of interactions between
cooperators and defectors and both the fraction of interface and active
sites increase. For the region around $S=0.65$, the only difference 
in the ranking of payoffs is that we now have $p(K_1^1)>p(K_0^2)$, 
while for $S=0.5$ it was $p(K_1^1)<p(K_0^2)$. Interestingly, although a cooperator with
a single cooperative neighbor fares better for $S=0.65$ than for 0.5, when
compared with a defector with two cooperating neighbors, the density of 
cooperators is strongly reduced when compared with the neighboring regions. 

If one ranks all values of $p(K_{\theta}^n)$, for the values of $T$ and $S$ considered
in Fig.~\ref{fig.rho1.snapshots}, as $S$ increases, the unique modification is that
$p(K_0^2)$ moves further down in the payoff ranking. For $S=0.8$, for
example, $p(K_0^2)$ and $p(K_1^0)$ switch places (when compared with $S=0.65$) and $p(K_0^2)<p(K_1^0)$.
A sublattice of cooperators (or, equivalently, defectors) separated by every other site is 
stable, while the intermediate
sites may flip from one to the other. For a random initial state, the lattice will be
populated with small patches of such a stable structure. 
In the PD region, cooperation is
sustained by compact groups of cooperators while they decrease in compactness as $S$
becomes larger, accompanied by an increase of cooperator-defector interfaces, 
since unilateral cooperation becomes worthwhile. It is important to emphasize that
although it seems at first that the increase of $S$ would enhance 
cooperation in a population, what it indeed promotes is a continued interaction between cooperators
and defectors, since the punishment for being exploited decreases. 
\begin{figure*}[bth]
\centering
\includegraphics[width=8cm]{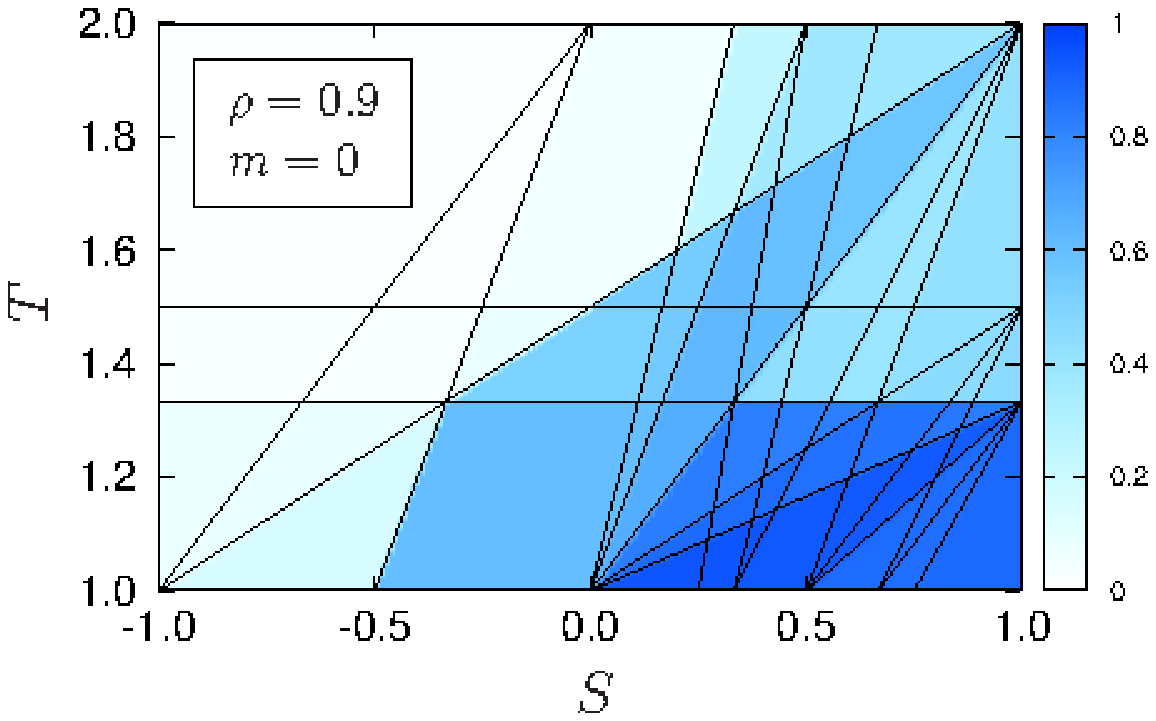}
\includegraphics[width=8cm]{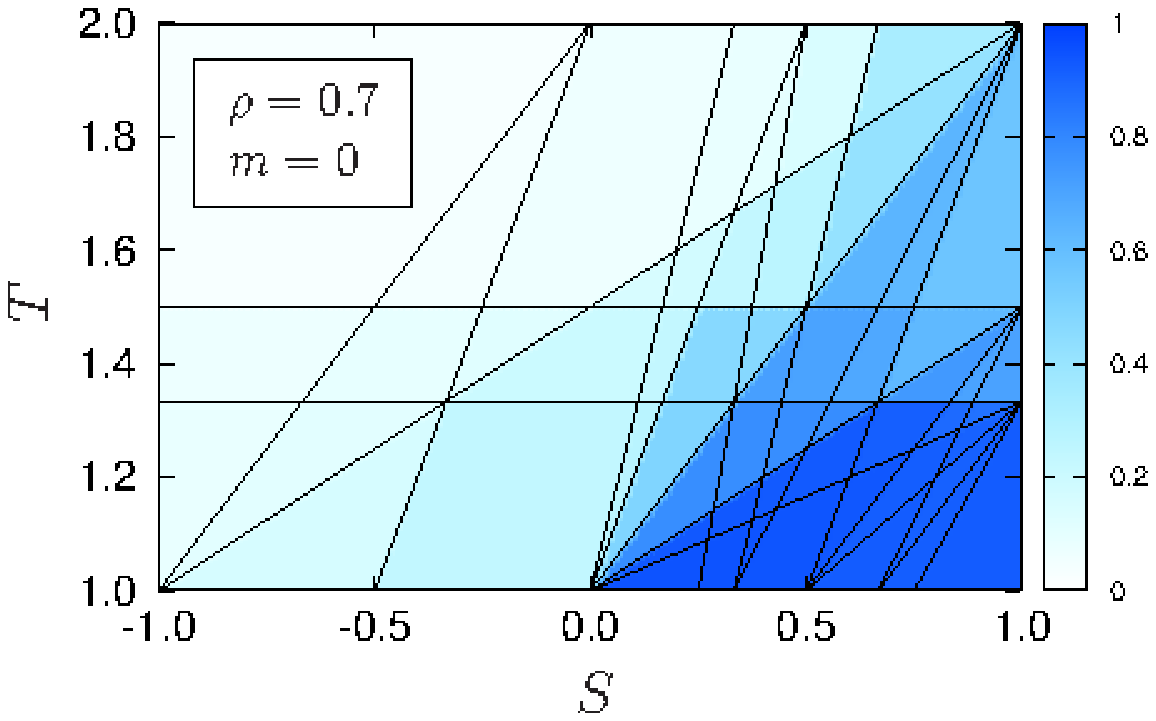}

\vspace{-8mm}
\includegraphics[width=8cm]{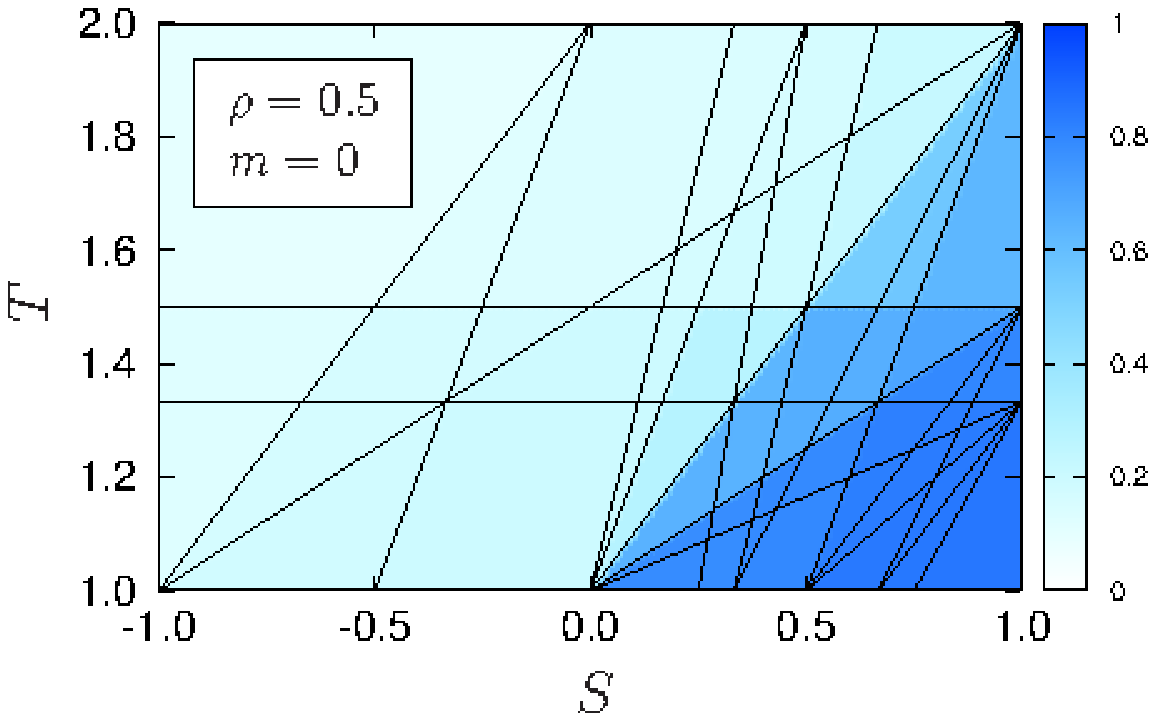}
\includegraphics[width=8cm]{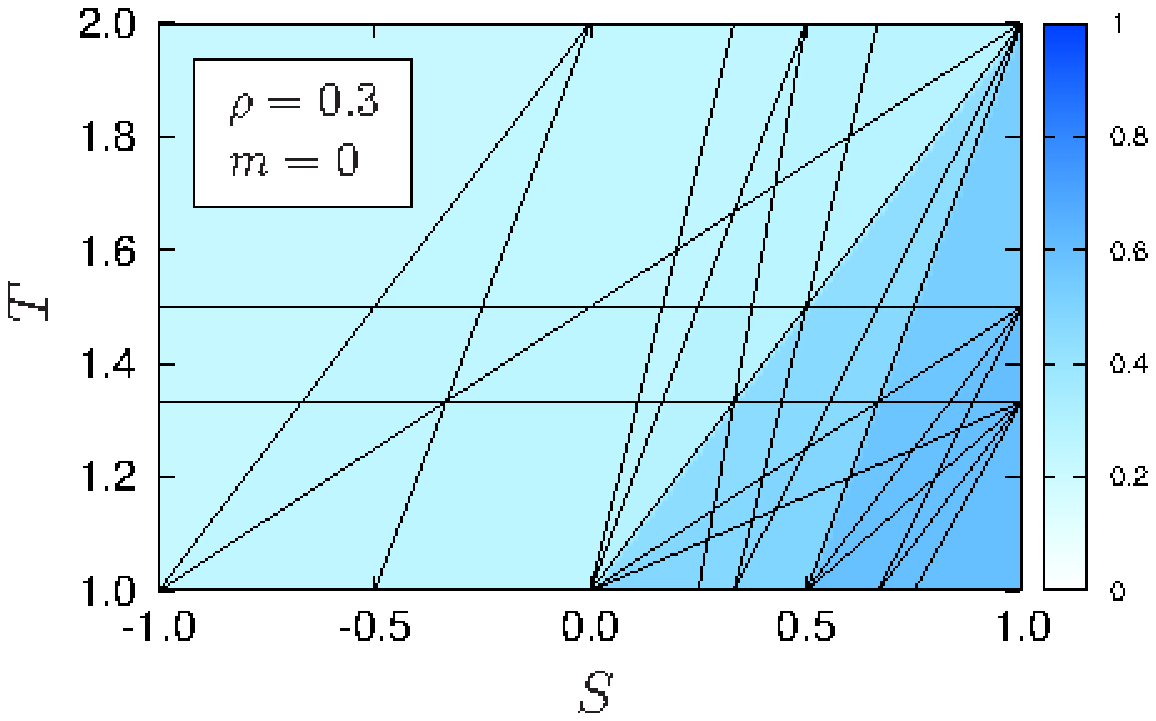}
\caption{Phase diagram for diluted lattices without mobility ($m=0$)
and several densities $\rho$. 
The color code indicates the level of cooperation $\rhoc/\rho$.
The solid lines represent the functions $f_{n_0n_1e_0e_1}$ which
 delimit different phases. Although the levels of cooperation are quite similar
to those of Fig.~\ref{fig.phase_diagram}, a few more lines (and, consequently, a large
number of new phases) are introduced due to dilution. This is the case of the four lines that cross
at $(S,T)=(1,3/2)$ and another one that passes through the point $(1,3)$. In the bottom figures, 
for $\rho=0.3$ and 0.5,
cooperation is sustained, even if at low levels, in all regions.}
\label{fig.diagram_empty_0.9}
\end{figure*}
In fact, one may
introduce a measure of such exploitation as the relative number of CD pairs 
($\rhoper$), also related
to the total perimeter of cooperator clusters and shown in Fig.~\ref{fig.rho1}, bottom panel. 
These interfaces can also be
directly observed in the snapshots of Fig.~\ref{fig.rho1.snapshots}, in the form of
checkerboard-like regions in which large groups of cooperators exist with nested
defectors. Notice
that although $\rhoc$ is not monotonic in $S$, $\rhoper$
is an almost monotonically increasing function of $S$ (the region $0.6<\rho<0.7$ is
very particular: besides the strong depression in the amount of cooperators, it also
breaks the monotonicity, as a function of $S$, of both $\rhoper$ and the fraction of
strategy switching, i.e. active, sites, $\rhoa$). 

Interestingly, the elongated structures of cooperators in the SD game, observed in the rightmost snapshot 
of Fig.~\ref{fig.rho1.snapshots}, are similar to those observed in Ref.~\cite{HaDo04} although
the dynamics and the parameters are not the same, indicating that the results found here for
a specific dynamical rule may be more generally valid. In addition, those dendritric structures are
but one way of creating large interface structures, alternatives being isolated cooperators or checkerboard-like
groups~\cite{SzSzVaHa10}.

\subsection{Diluted lattices ($\rho<1$) without mobility ($m=0$)}

Disorder may be included in these games in several different
ways, for example, as site~\cite{VaAr01} or bond~\cite{LiLiTiSh10} 
dilution. We consider here the former, once the mobility mechanism
that we will later use is dependent on the existence of empty sites. In Ref.~\cite{VaAr01}
we have seen that, for the weak version of the PD game, 
a small amount of disorder gives rise to pinning points that
prevent the strategy switching waves from traversing the system.
Indeed, groups of cooperators can be shielded by empty sites, 
what could be interpreted as natural landscape defenses, and keep their strategy
for long intervals of time. These long lasting strategies may be observed, for
example, by measuring the persistence function, the fraction of agents that did not
switch strategy since $t=0$. The existence of an asymptotic zero persistence
has been shown~\cite{VaAr01} to be related to the existence of an expressive number of active 
sites (those that changed strategy since the last time step) at larger
densities, while for smaller ones the persistence attains a finite plateau and
there is a vanishing number of such active sites. In particular, there is an 
optimal intermediate density at which cooperators have a maximum population, what
remains valid even when the imitating updating dynamics is stochastic~\cite{WaSzPe12}.
In this last case, in which the optimal cooperative state is closely related to the 
percolation threshold~\cite{WaSzPe12}, the existence of fractal clusters at the
threshold seems to be important as neither disconnected nor compact clusters are present
that help defectors to invade and exploit cooperator communities.

\begin{figure}[htb]
\includegraphics[width=8.5cm]{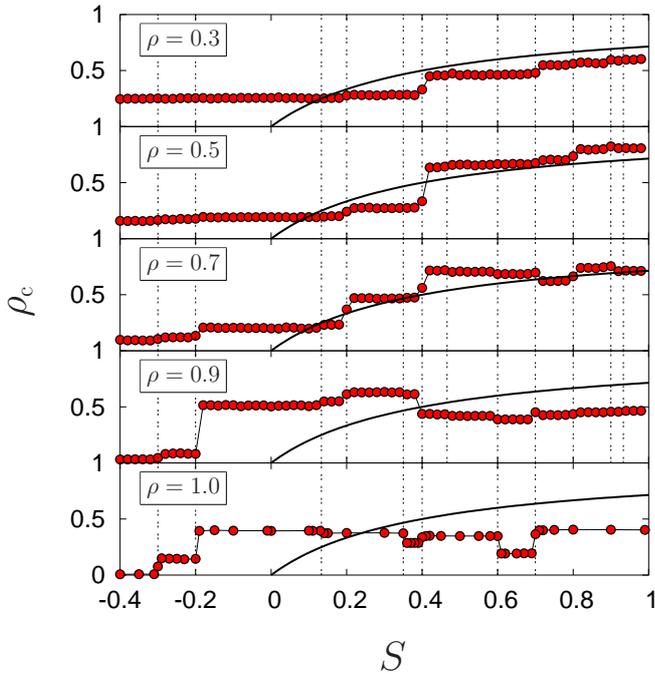}
\caption{Average fraction of cooperating individuals $\rhoc$
versus $S$ ($T=1.4$, $R=1$ and $P=0$) for different values
of the density $\rho$ without mobility ($m=0$). The panel at
the bottom shows, as vertical lines, the transition points for 
$\rho=1$: $S=-0.3$, $-0.2$, 2/15, 0.35, 0.4, 0.6  and 0.7.
In the remaining panels, in which $\rho<1$, new transitions appear at
$S=0.2$, 7/15, 0.8, 0.9 and 14/15. Interestingly, besides the transition
at $S=0.4$ representing a strong change in $\rhoc$, whether the jump is 
upward or downward depends on the value of $\rho$. 
Also shown (curved line) is the mean field result. Notice that
for large values of $S$, small and large densities fare worse than the mean field and
cooperators perform better only for intermediate densities in the presence of 
spatial correlations.}
\label{fig.m0}
\end{figure}

We here extend the results of Ref.~\cite{VaAr01} for other 
values of $S$ (see also Ref.~\cite{SiFoVaAr09}),  
and explore its microscopic interpretation.
 The introduction of empty sites changes the phase diagram by allowing 
new configurations of local structures. Therefore, the phase diagram will be composed by the
lines that were already present in the case without empty sites, Fig.~\ref{fig.phase_diagram}, 
plus a few more.
Besides such new phases, the amount of cooperation will also depend on the total 
density $\rho$~\cite{VaAr01}. 

Extending the notation introduced earlier, the local neighborhoods shall be denoted 
by $K_{\theta}^{ne}$, where $\theta \in \{0,1\}$ describes the occupation of the
center cell,  $e \in \{0,1,2,3,4\}$ gives the total number of empty sites in the local 
neighborhood and $n \in \{0,1,2,3,4\}$ ($n \leq 4-e$ and $n_0 \neq 0$) gives the total
number of cooperators in the local neighborhood. Now, the number of defectors is 
given by $4-n-e$. In this way, we have $K^{n0}_{\theta} \equiv K^n_{\theta}$. In addition to the 
functions given in Eq.~(\ref{eq.diag}), the following functions which compare the local 
neighborhoods $K^{n_0e_0}_0$ and $K^{n_1e_1}_1$, with $e_1 \neq 0$, should also be taken into 
account
\begin{equation}
f_{n_0n_1e_0e_1}=\frac{n_1 R+[4-(n_1+e_1)]S-[4-(n_0+e_0)]P}{n_0}.
\label{eq.diag_empty}
\end{equation}
 Using the above notation, $f_{n_0n_100}\equiv f_{n_0n_1}$. Not all of these 
functions are used, since many give conditions in the region $T
\leq R$ (valid for other games).

\begin{figure}[htb]
\includegraphics[width=8.5cm]{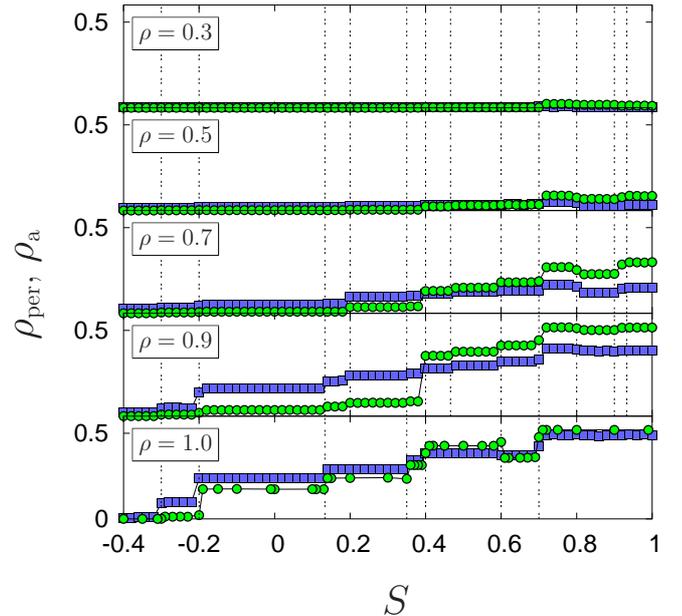}
\caption{Average fraction $\rhoa$ of active (green/light grey circles) and
pairs of opposite strategies (blue/dark grey squares), $\rhoper$,
versus $S$ ($T=1.4$, $R=1$ and $P=0$) for different values
of the density $\rho$ without mobility ($m=0$). The transition
lines follow Fig.~\ref{fig.m0}.}
\label{fig.m0_per}
\end{figure}

\begin{figure}[tbh] 
\includegraphics[width=3.5cm]{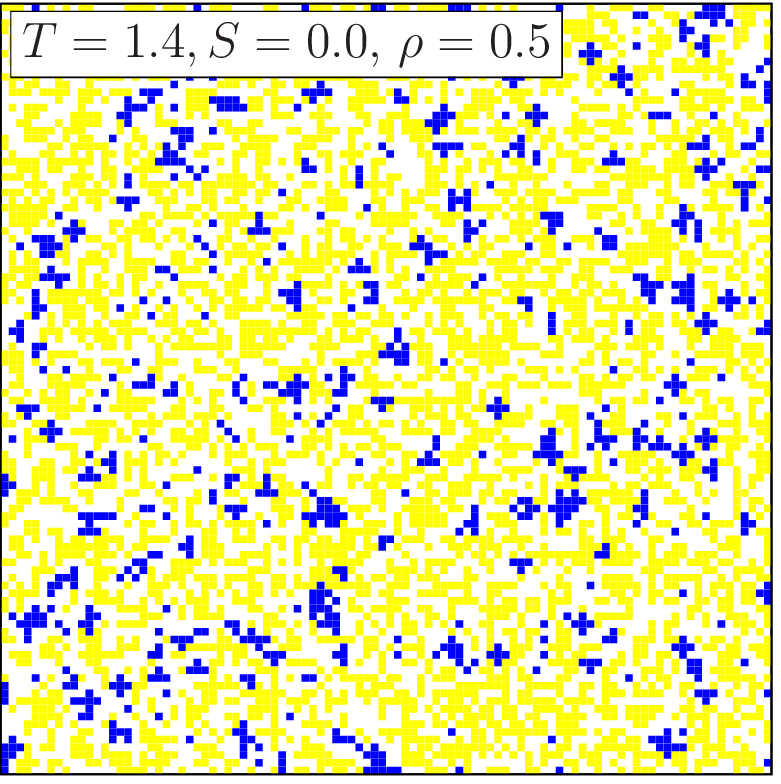}
\includegraphics[width=3.5cm]{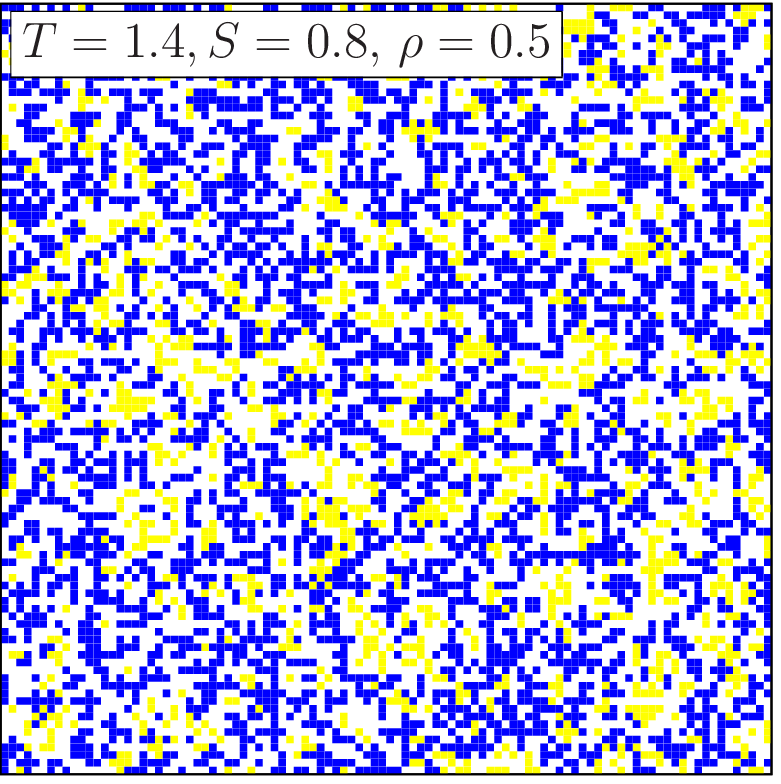} 

\includegraphics[width=3.5cm]{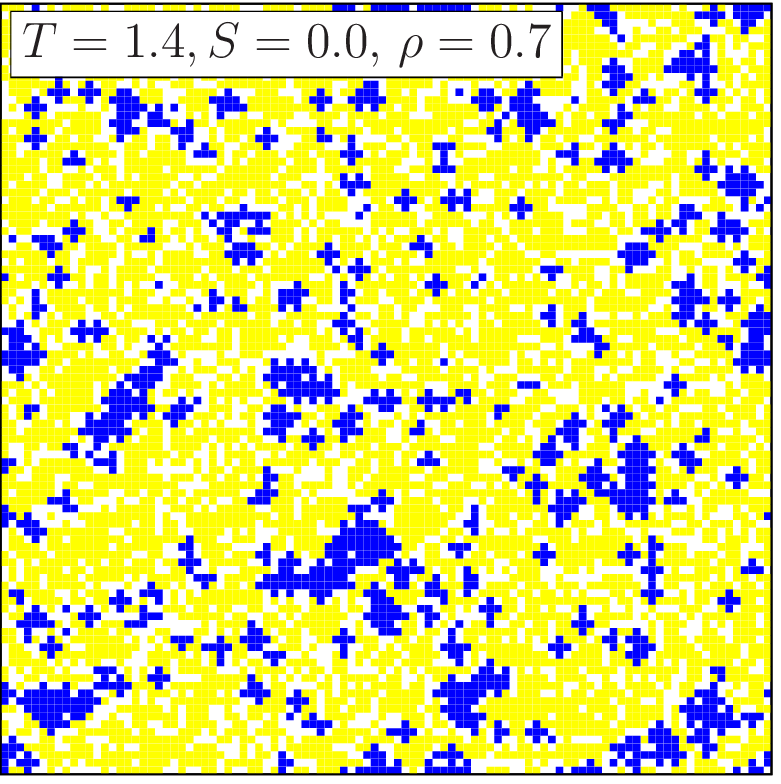}
\includegraphics[width=3.5cm]{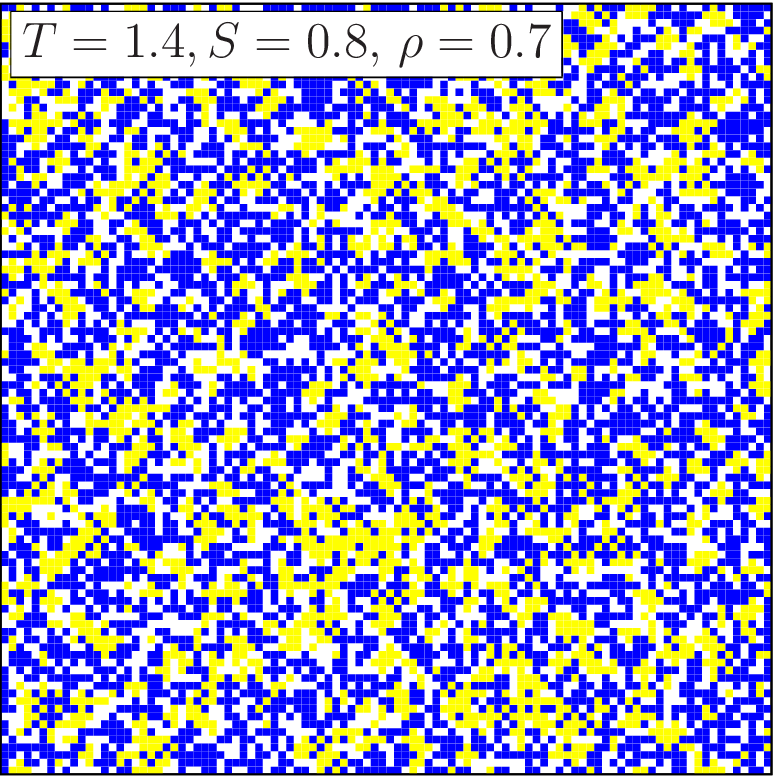} 

\includegraphics[width=3.5cm]{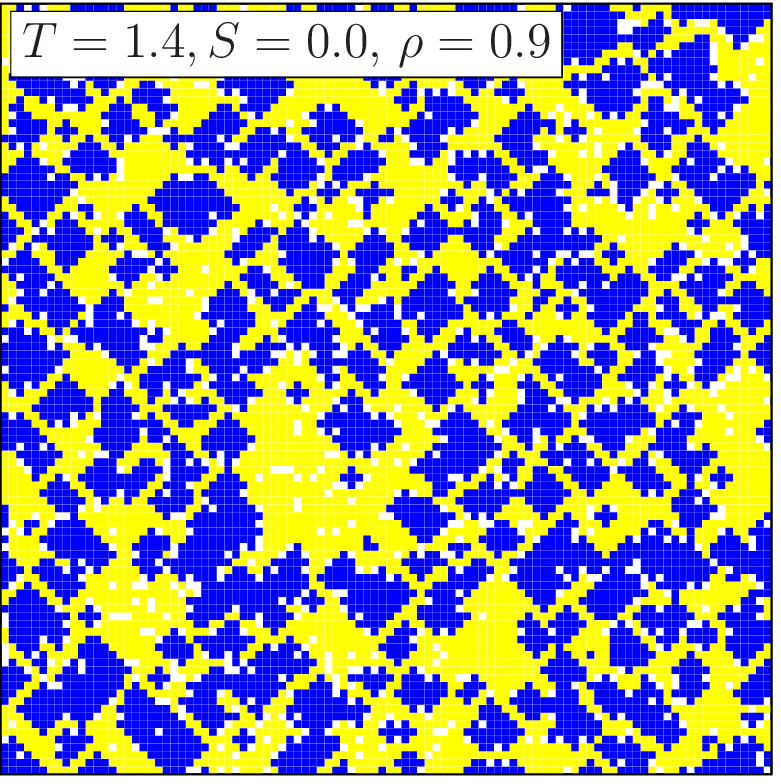}
\includegraphics[width=3.5cm]{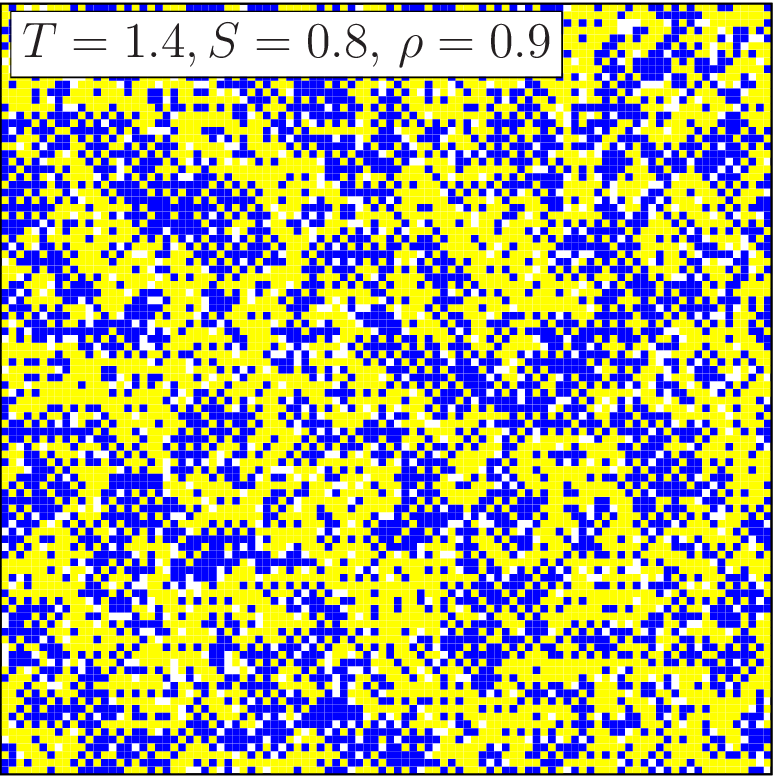} 
\caption{Snapshots for $m=0$, after $10^4$ MCS, showing typical configurations for several values
of $S$ with $T=1.4$ and densities both below and above the site random percolation 
threshold ($\rho\simeq 0.59$). 
Blue/yellow (black/grey) sites represent cooperators/defectors, as in Fig.~\ref{fig.rho1.snapshots},
while white ones are empty sites. Notice that for a given density, the empty sites in these
snapshots are in the same position.}
\label{fig.rho0.5.snapshots}
\end{figure}

The values of $(n_0,n_1,e_0,e_1)$ that contribute to the diagram in the range $T>R$ and
which lead to functions different from the ones listed for the case $\rho=1$ are: 
$(2,3,0,1)$, $(2,2,0,1)$, $(2,1,0,1)$, $(2,0,0,1)$, 
$(1,3,0,1)$, $(1,2,0,1)$, $(1,1,0,1)$, $(1,0,0,1)$, 
$(1,0,0,2)$, 
$(1,2,0,2) \equiv (2,4,0,0)$ and $(1,1,0,2) \equiv (2,2,0,0)$.
It should be noted that this diagram is only valid for the case $P=0$, because in this
 case a D with a D neighbor is equivalent to a D with an empty neighbor. If $P\neq 0$, 
then these two configurations are not the same, what will give rise to further
phase separating lines.
Fig.~\ref{fig.diagram_empty_0.9} shows the diagrams
for several values of $\rho$. For $\rho=0.9$, it is not very different from the
full $\rho=1$ case, apart from an intensification of cooperation in the lower
right corner of the figure, what is consistent with the results of Refs.~\cite{VaAr01,WaSzPe12}
that showed that a small amount of quenched dilution is an enhancement factor for cooperation as it
prevents defectors from invading cooperator clusters.
Stronger deviations are observed for $\rho=0.5$: although
presenting cooperation in the whole region shown, the phases that presented 
cooperation previously now have a smaller density of cooperators. 
For small densities, mainly below the percolation threshold, the fate of
isolated clusters only depends on their initial composition of Cs and Ds.
Thus, regions that were previously unable to sustain cooperation now have small
but finite fractions of cooperators (e.g., in the left top corner of the phase
diagrams in Fig.~\ref{fig.diagram_empty_0.9}). For even smaller densities, the
final configuration differs little from the initial one, $\rhoc$ tends to 1/2 and
the phase diagram becomes homogeneous, independent of $S$ and $T$.

An example of
the sudden transitions occurring at the delimiting lines between phases is shown in Fig.~\ref{fig.m0}
for a cut at $T=1.4$ for several densities~\cite{SiFoVaAr09}. Again, the
transitions can be observed as jumps in the value of $\rhoc$ at the
specified points.
The most notable transition occurs at $S=0.4$:
depending on the total density $\rho$, the fraction of cooperators
may either jump upwards or downwards, and the change is much more
pronounced than for $\rho=1$. Cooperation is enhanced at intermediate
densities and may even fare better than the mean field (e.g., for
$\rho=0.5$ and 0.7 in the figure). The structures formed by cooperators
also follow the overall pattern observed for $\rho=1$ and a few examples are shown
in Fig.~\ref{fig.rho0.5.snapshots}: compact groups in the PD game (left column)
and dendritic or checkerboard like in the SD game. For $S=0$ (left column), although the ever
present small cooperator clusters start to increase in size after the
percolation threshold, only well above this transition point do they occupy a large
fraction of the network. The optimal density for cooperators is not that
high when stochastic rules are used, being shifted towards the threshold~\cite{WaSzPe12}.
For $S=0.8$ (right column), on the other hand, cooperators group themselves
into dendritic or checkerboard structures (or stay isolated).

The density of cooperators is monotonic in $S$ only for very low
densities, Fig.~\ref{fig.m0}, while for large $S$ (most notably for
$S>0.4$), $\rhoc$ tends to decrease and becomes non monotonic. More
information can be obtained by measuring the fraction of
active sites, $\rhoa$, and the fraction of pairs of different 
strategies, $\rhoper$, as shown in Fig.~\ref{fig.m0_per}.
For low densities, $\rho=0.3$ and 0.5 in Fig.~\ref{fig.m0_per},
the configuration is almost frozen and both quantities
are very close to zero. Otherwise,  they present a tendency to increase 
with $S$ (albeit exceptional intervals are still present). 
Interestingly, although
there seems to be a correlation between the two parameters for all
densities, deviations are stronger close to the optimum
density (see the case of $\rho=0.9$ in Fig.~\ref{fig.m0_per}).

\subsection{Diluted lattices ($\rho<1$) with mobility ($m\neq0$)}

\begin{figure*}[htb]
\includegraphics[width=8cm]{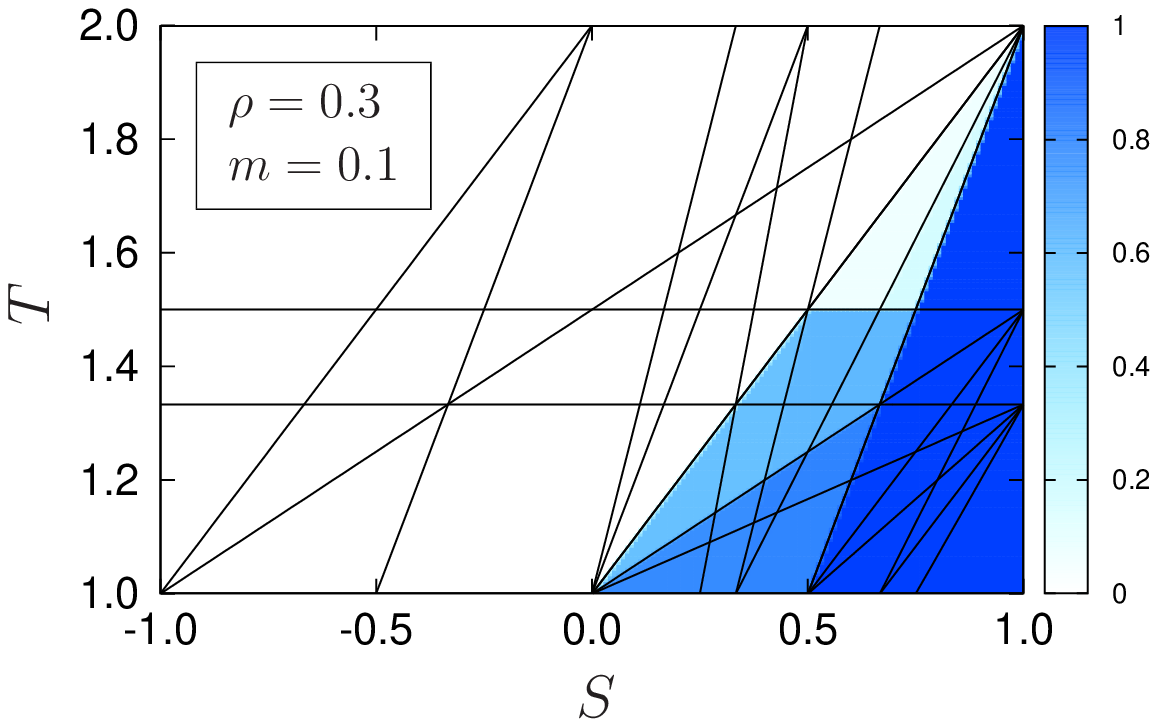}
\includegraphics[width=8cm]{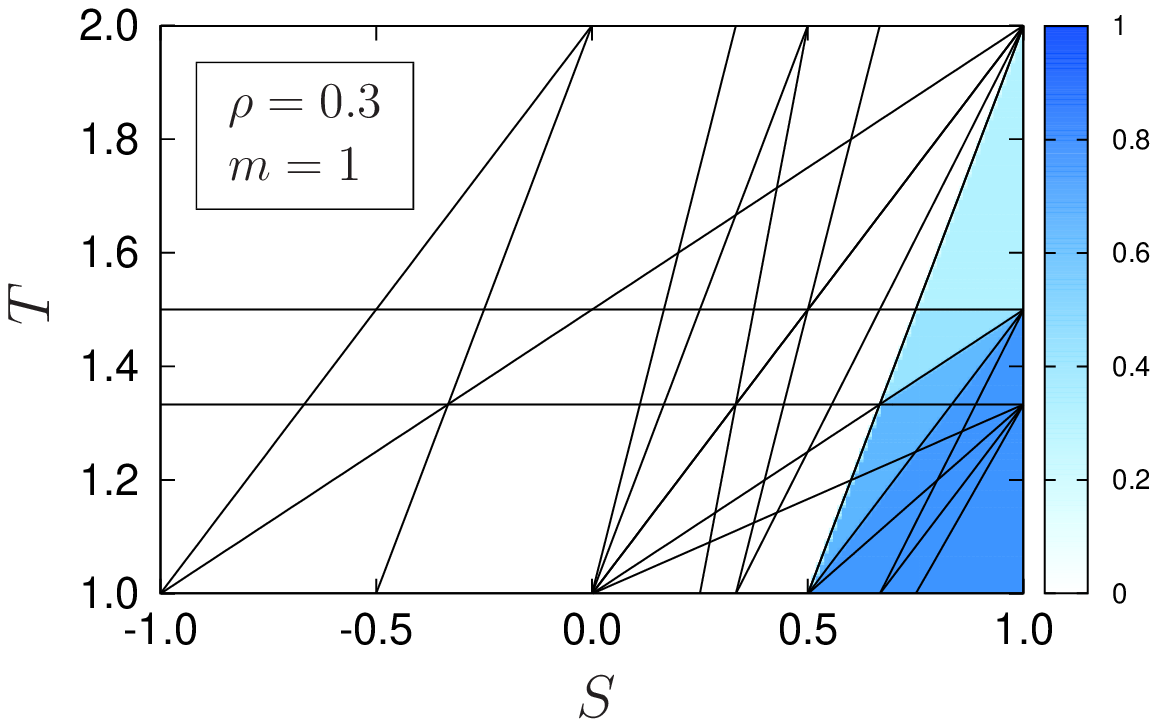}

\vspace{-8mm}
\includegraphics[width=8cm]{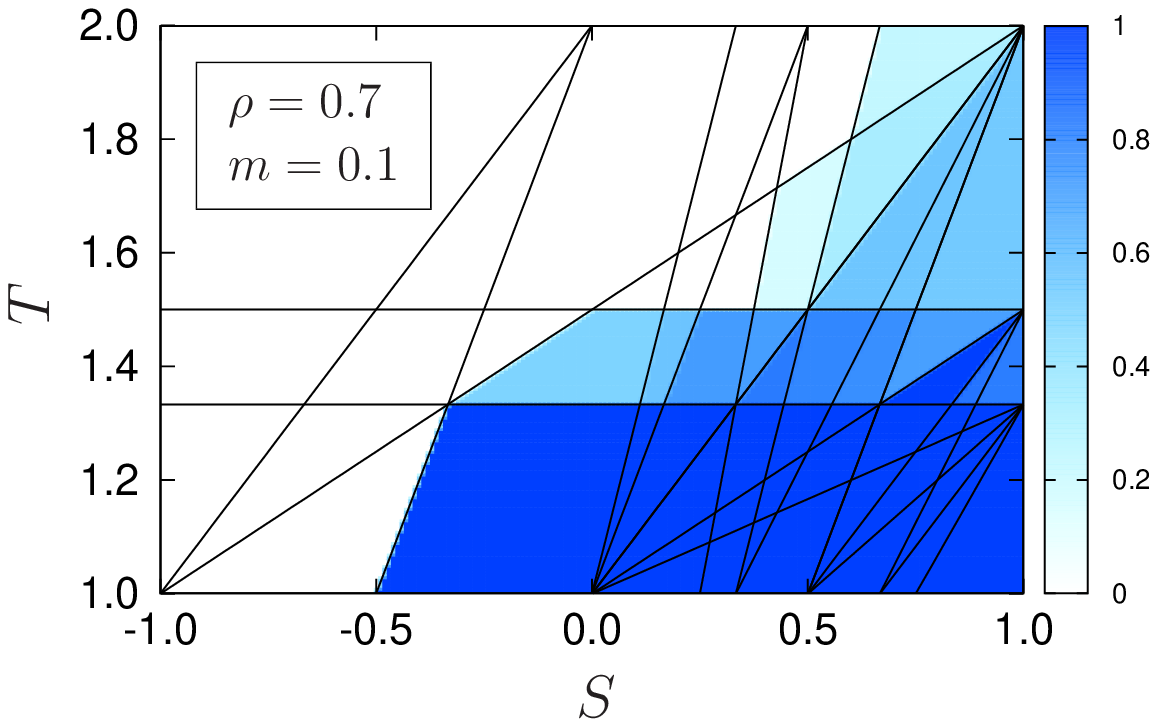}
\includegraphics[width=8cm]{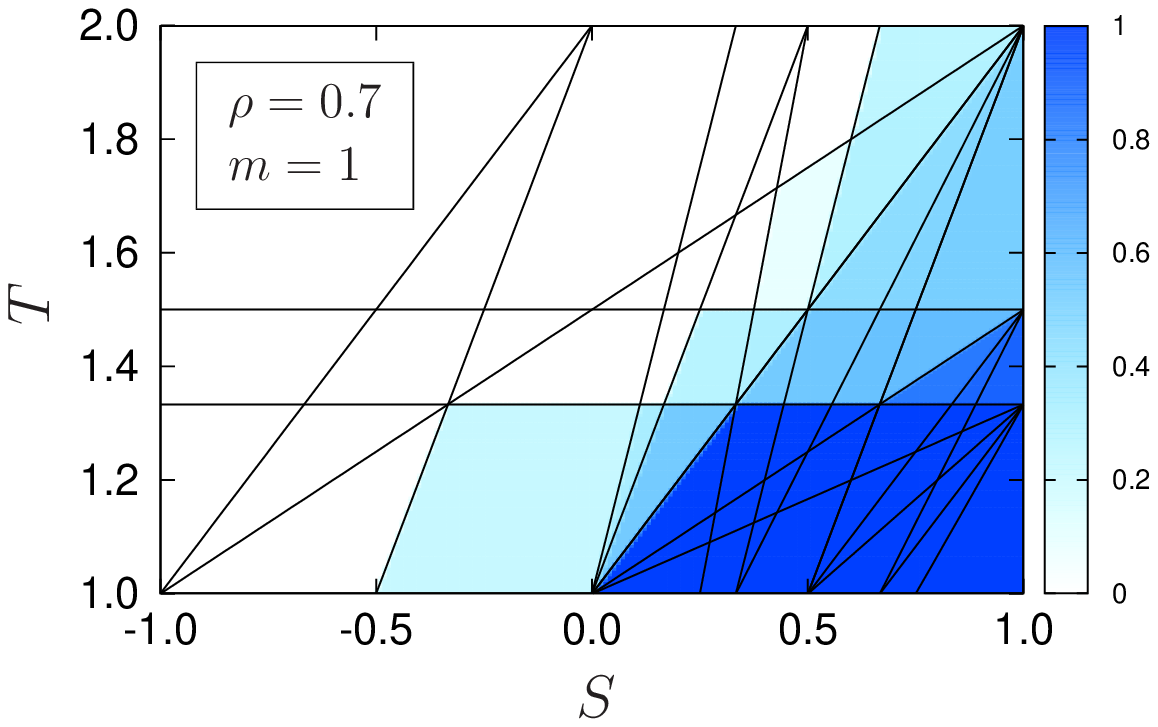}
\caption{Phase diagrams for the case in which the diffusive step is taken
after the offspring generation (COD). The smaller mobility has more
cooperative regions in the phase diagram (blue) and an overall
higher level of cooperation in each region. Notice also that the PD game
does not present cooperation at low densities.}
\label{fig.phase_diagram_COD}
\end{figure*}

\begin{figure*}[tbh] 
\includegraphics[width=3.9cm]{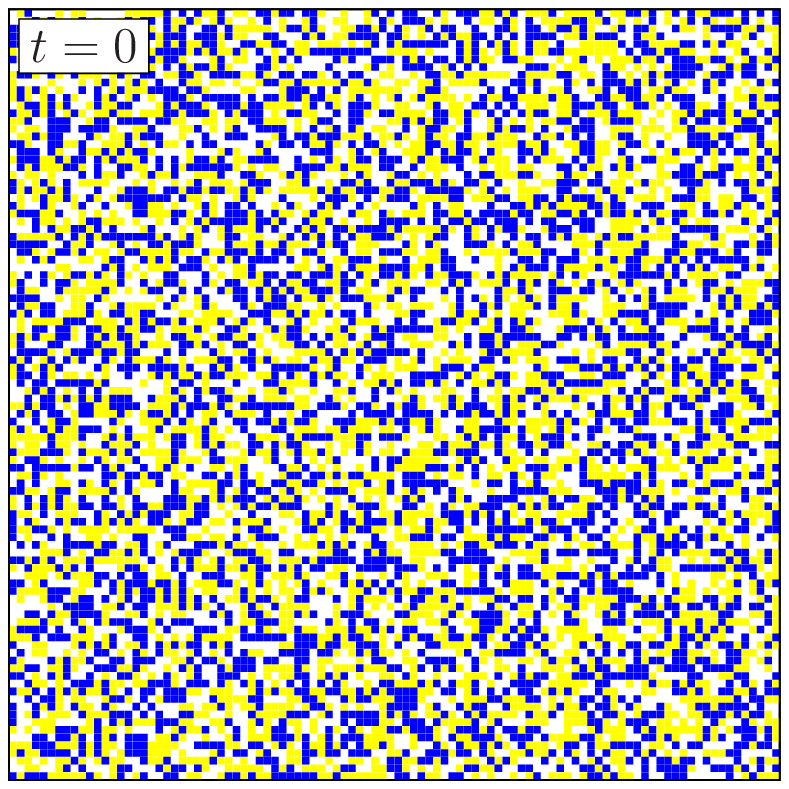}
\includegraphics[width=3.9cm]{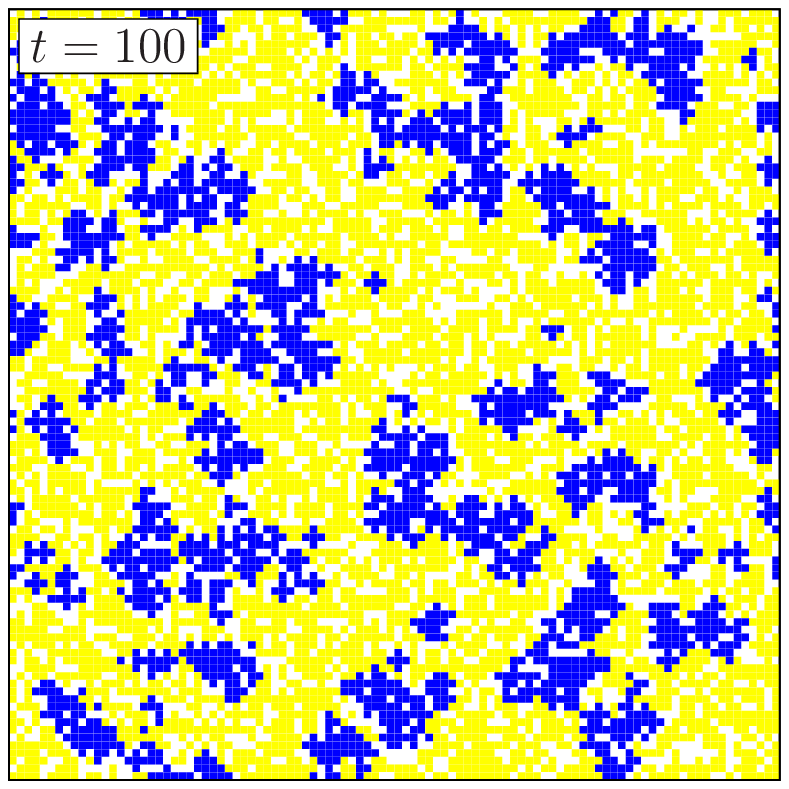} 
\includegraphics[width=3.9cm]{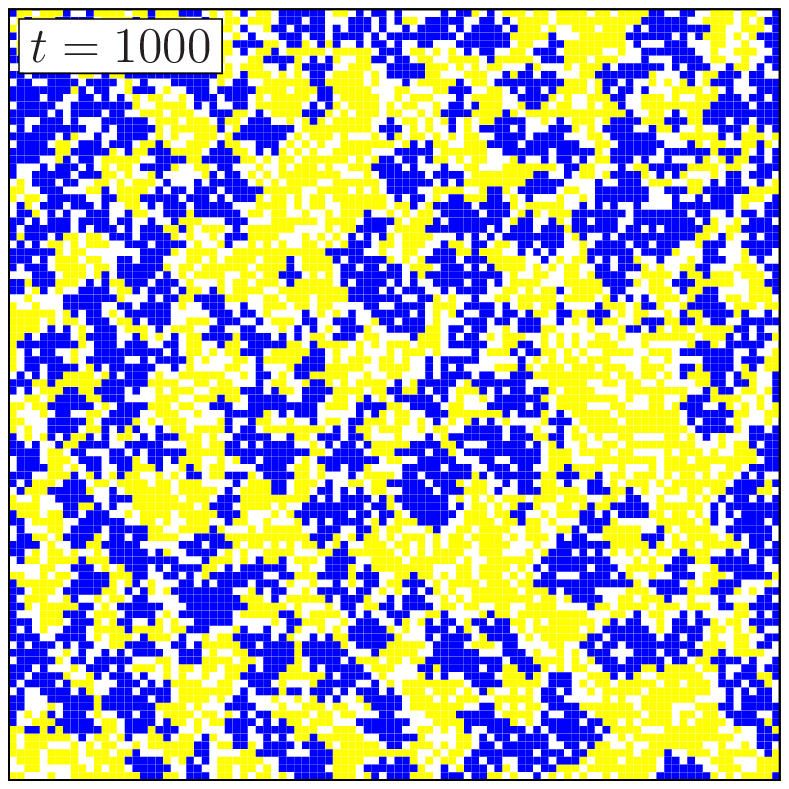}
\includegraphics[width=3.9cm]{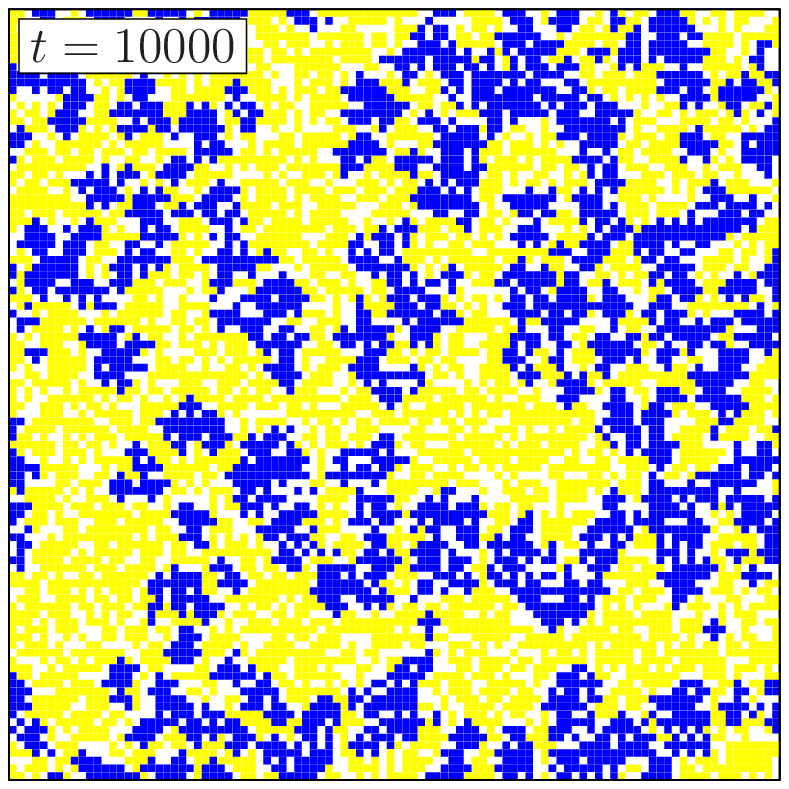}

\includegraphics[width=3.9cm]{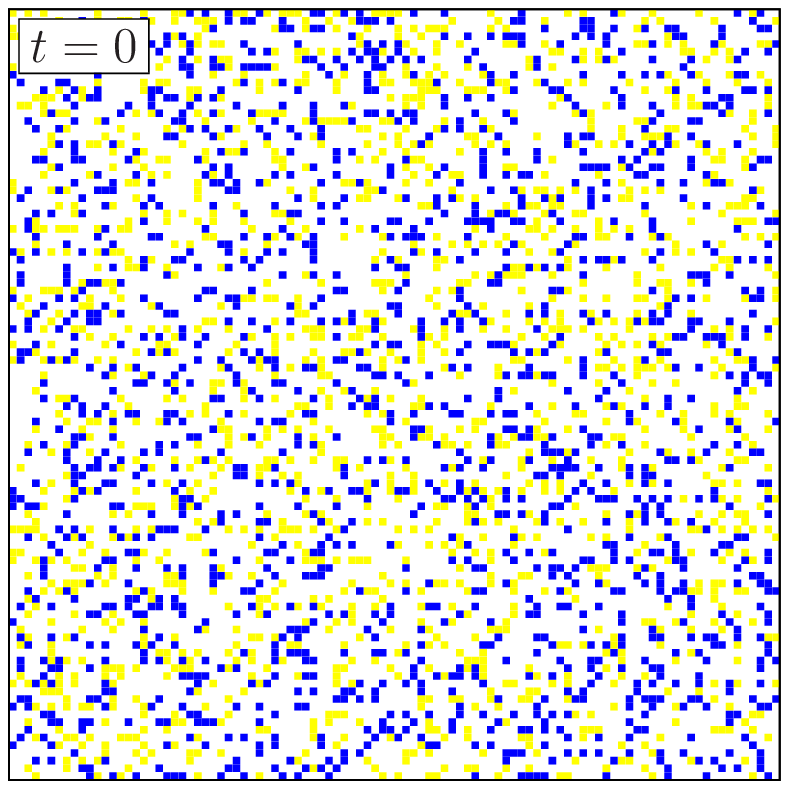}
\includegraphics[width=3.9cm]{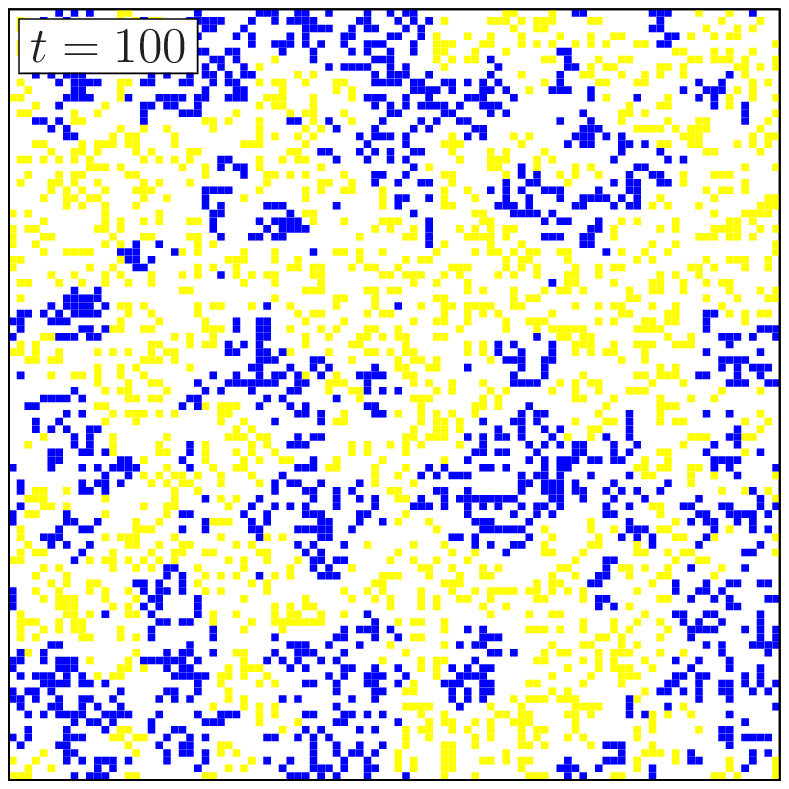} 
\includegraphics[width=3.9cm]{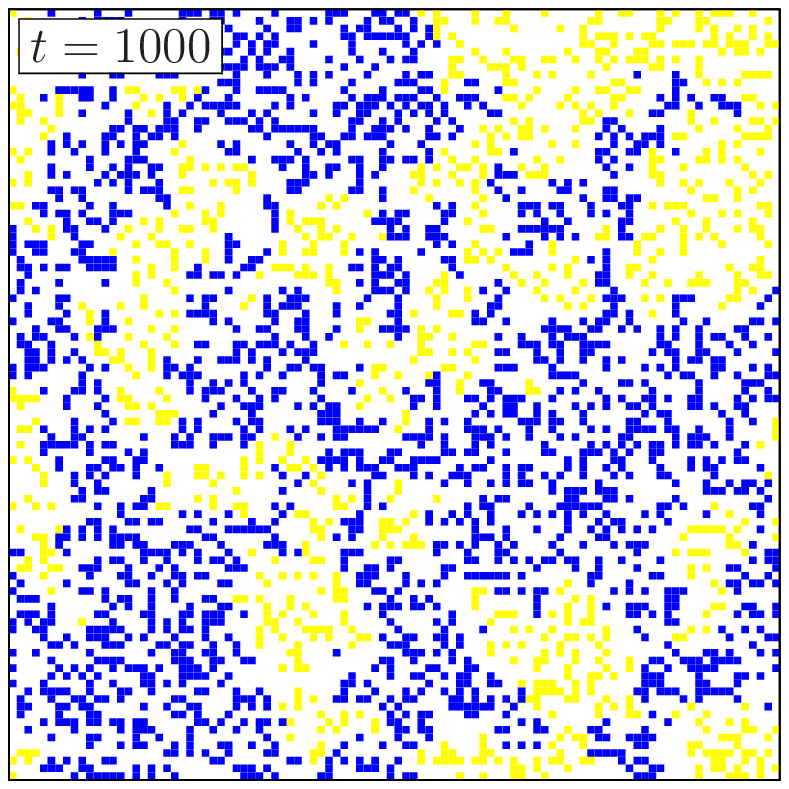}
\includegraphics[width=3.9cm]{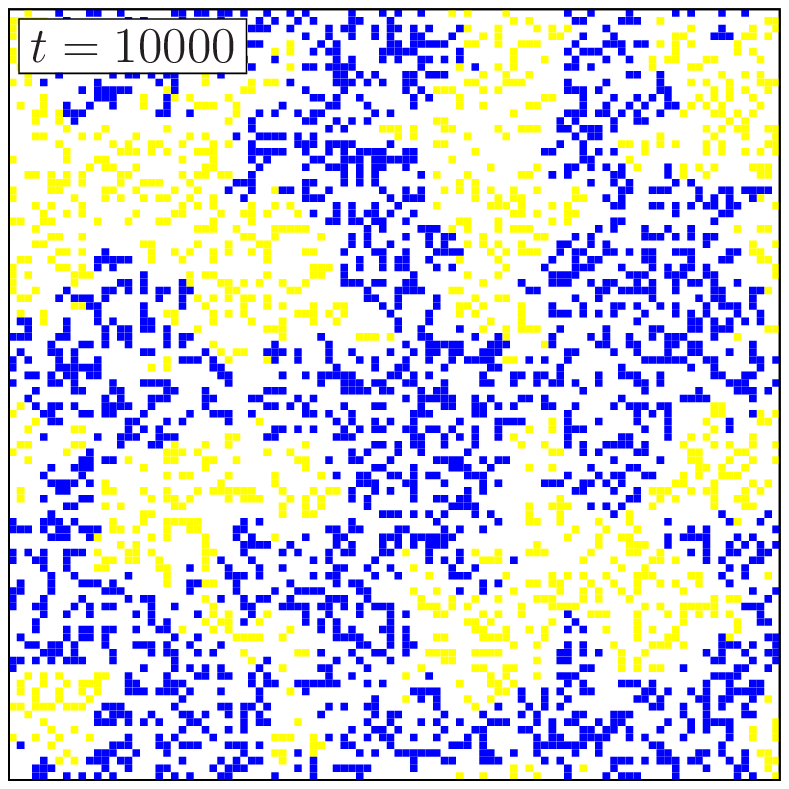} 
\caption{Snapshots for $m=0.1$ and $T=1.4$  with COD dynamics for 
several times, showing typical configurations. The top row is for
$\rho=0.7$ and $S=0$, while the bottom row shows $\rho=0.3$ and $S=0.5$. 
Blue/yellow (black/grey) sites represent cooperators/defectors, 
as in Fig.~\ref{fig.rho1.snapshots}, while white ones are empty sites. The bottom
row, despite its small density (below the percolation threshold), has percolating 
defector-free regions. The upper row presents, on the other hand, cooperator-free
percolating regions.}
\label{fig.rho.COD.snapshots}
\end{figure*}

In the case $m=0$ studied in Ref.~\cite{VaAr01} ($T=1.4$ and $S=0$), 
and for the noiseless imitation rule considered here, the fraction of cooperators 
starts to increase again around the site random percolation threshold ($\rho\simeq 0.593$ 
in the square lattice). At low densities, cooperators persist on some isolated clusters because of
a favorable initial condition that allowed them to overcome defectors. In this regime, 
and depending on how mobility is implemented, defectors may act as free riders that 
eventually exploit the whole
system (remember that random diffusion is a disaggregating factor), leading to the extinction of
cooperators. As shown in Ref.~\cite{VaSiAr07}, there is a minimum density above which cooperators
are able to survive in the presence of mobility. Although in some cases such density
is above the percolation threshold, and cooperation seems to need an underlying percolating cluster
in order to be maintained when $m\neq 0$, cooperation may also resist below the 
percolation threshold. Indeed, with a smaller temptation, isolated groups of cooperators are less
predated by defectors and are able to survive.

Whether or not the phase diagram changes when random mobility is also taken into account depends
on the details of the diffusion. When the offspring step is performed before the
diffusion (COD -- combat-offspring-diffusion dynamics -- in the notation of Ref.~\cite{VaSiAr07}), 
no new transition appears, and the phase diagram has the same cross sections as those of the previous 
section, whatever the value of $m$, albeit with different fractions of $\rhoc$ in each phase. 
These diagrams are shown in Fig.~\ref{fig.phase_diagram_COD}. Notice that many regions now
are fully dominated by defectors. In general, low mobility (left column) is more favorable to 
cooperation: besides being present in more regions of the phase diagram, the fraction of cooperators is 
also higher. However, the SD game benefits much more from mobility than the PD,  as most of the
shaded regions are located for $S>0$. The latter, in particular, only presents a finite fraction of
cooperators for high densities (see also Ref.~\cite{SiFoVaAr09}). Although it is difficult to summarize
its general behavior, the SD game fares better at intermediate densities.
In the snapshots of
Fig.~\ref{fig.rho.COD.snapshots} we observe how cooperators are able to build clusters even in 
the presence of random mobility. These clusters, however, are less compact than those for the case
without mobility, Fig.~\ref{fig.rho0.5.snapshots}. An interesting aspect in those snapshots is
the presence of defector and cooperator-free regions. In the upper row, for $\rho=0.7$ (above
the percolation threshold), cooperators aggregate in isolated domains while the defectors
percolate throughout the lattice. In the bottom row, on the other hand, although the connected
domains are smaller, a different kind of order can be observed: cooperators and vacant sites
form large regions free of defectors. These cooperators, due to the rattling aspect of diffusion,
are indeed correlated. Moreover, these defector-free regions now percolate. For this COD
dynamics, the agents do not carry any payoff with them, since the combat and offspring steps are performed 
in succession before diffusion. In other words, there is no memory of the previous location. 

Although evaporation is an important mechanism to decrease
cooperation (cooperators that move away from the surface of clusters tend to become
defectors), when its rate is not too large or if diffusion is prevented by geometric
hindrance at larger densities, clusters may be stable. If the  cluster surface is locally flat,
cooperators located at the surface make contact with three other cooperators and have high payoff. 
Moreover,
they also compare their payoff with interior cooperators whose payoff is even larger. These
regions are prone to cooperation and any defector that, after diffusion, gets in contact
with the cooperative surface will be assimilated. This is the basic growth mechanism in the COD
case.

\begin{figure*}[ht]
\includegraphics[width=8cm]{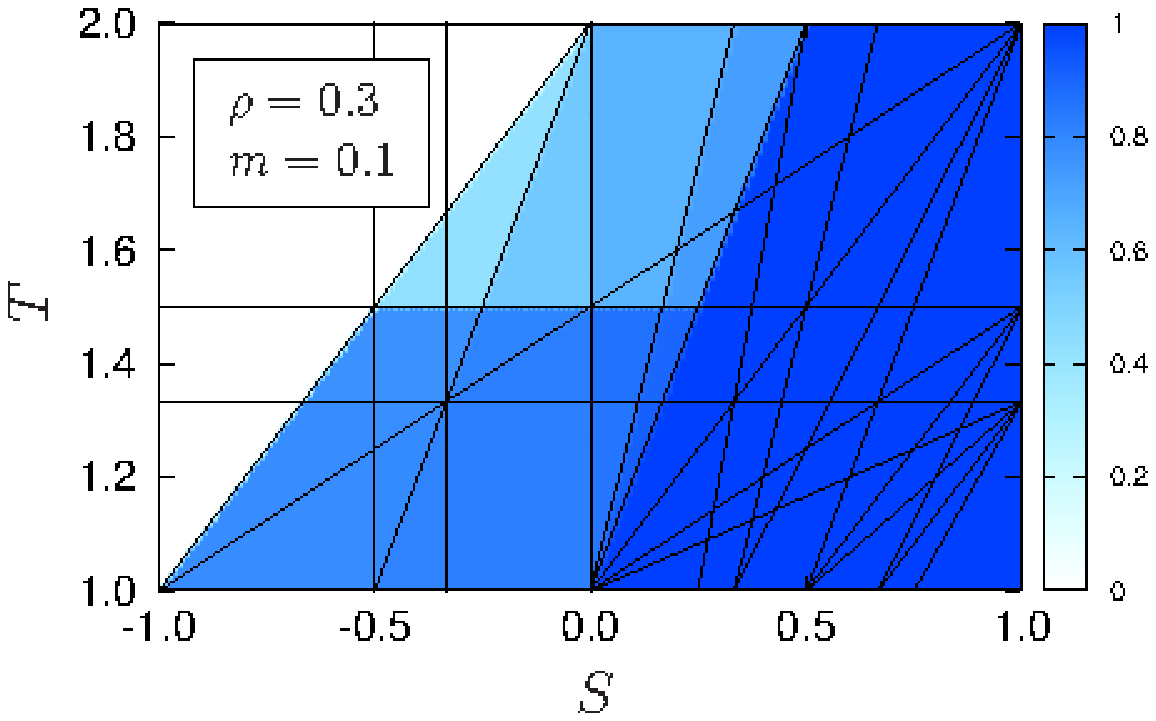}
\includegraphics[width=8cm]{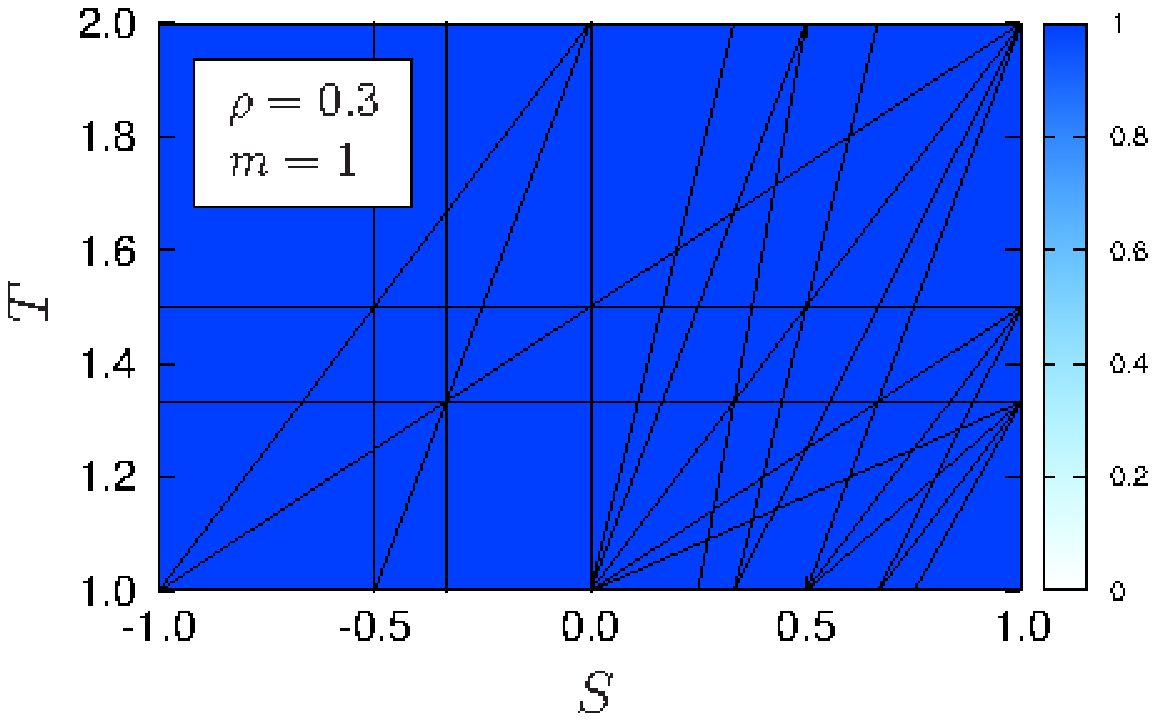}

\vspace{-8mm}
\includegraphics[width=8cm]{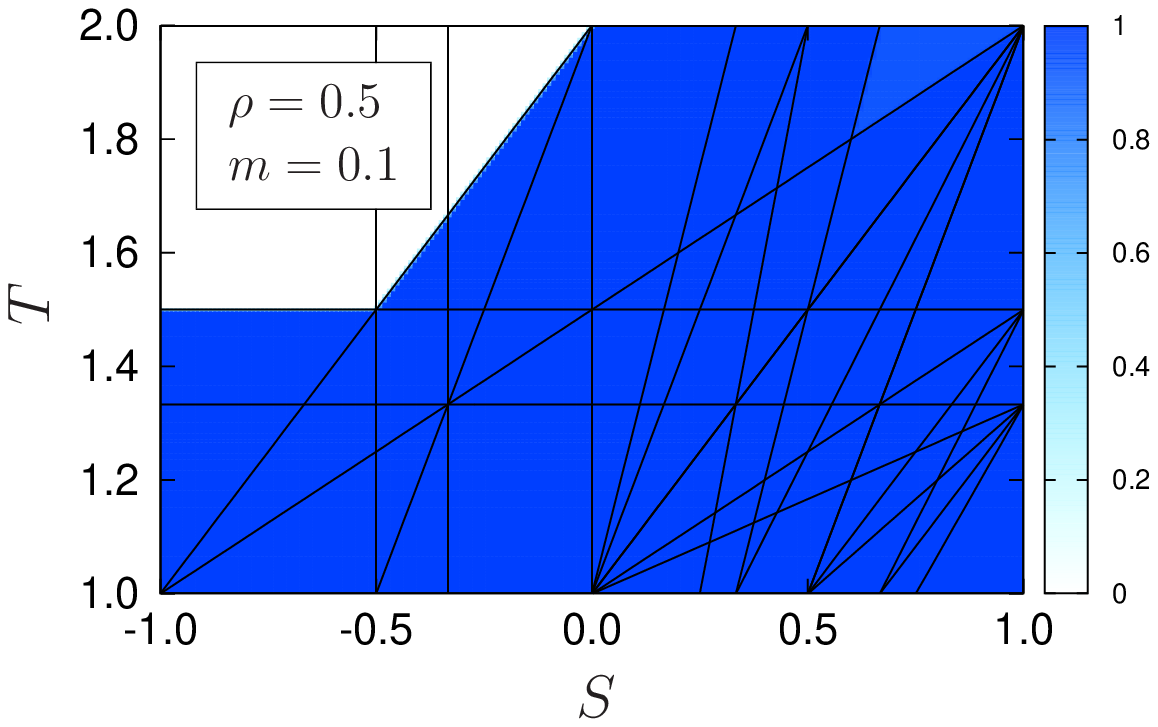}
\includegraphics[width=8cm]{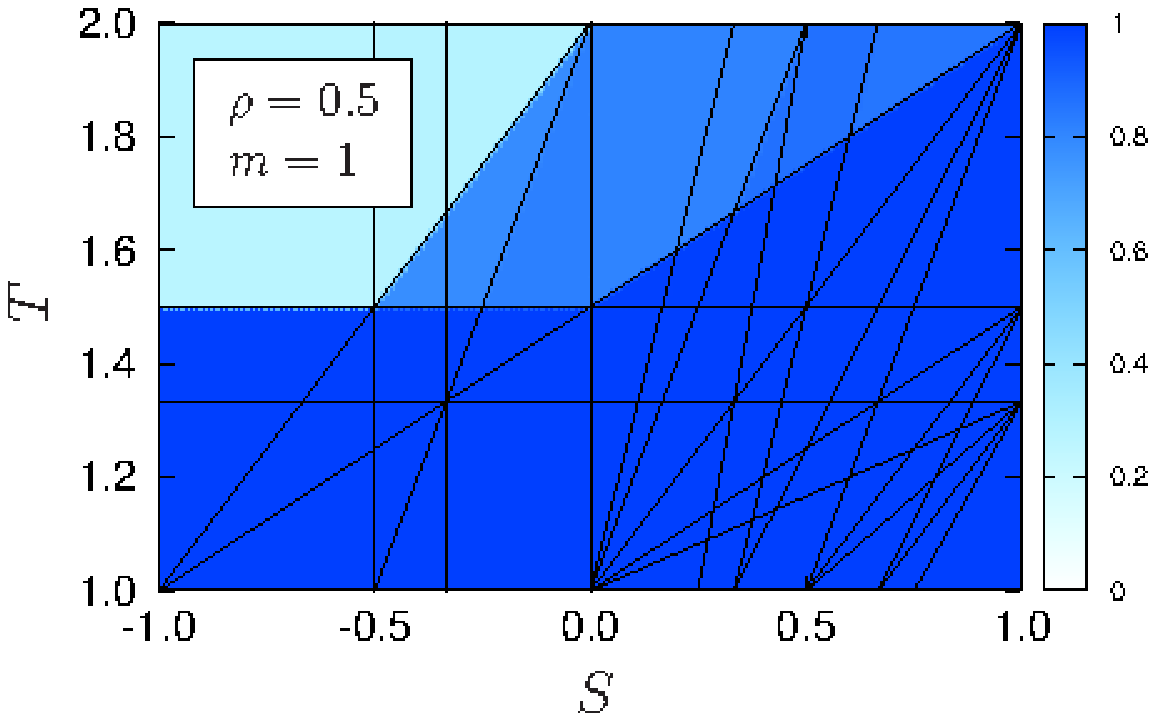}

\vspace{-8mm}
\includegraphics[width=8cm]{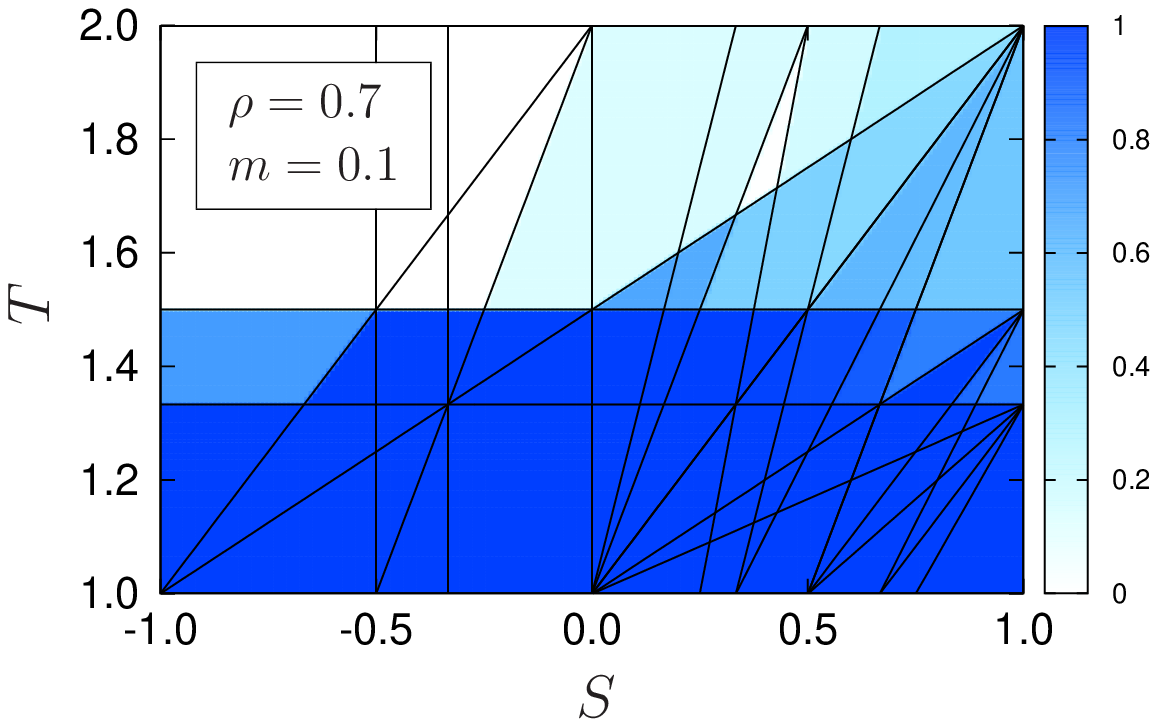}
\includegraphics[width=8cm]{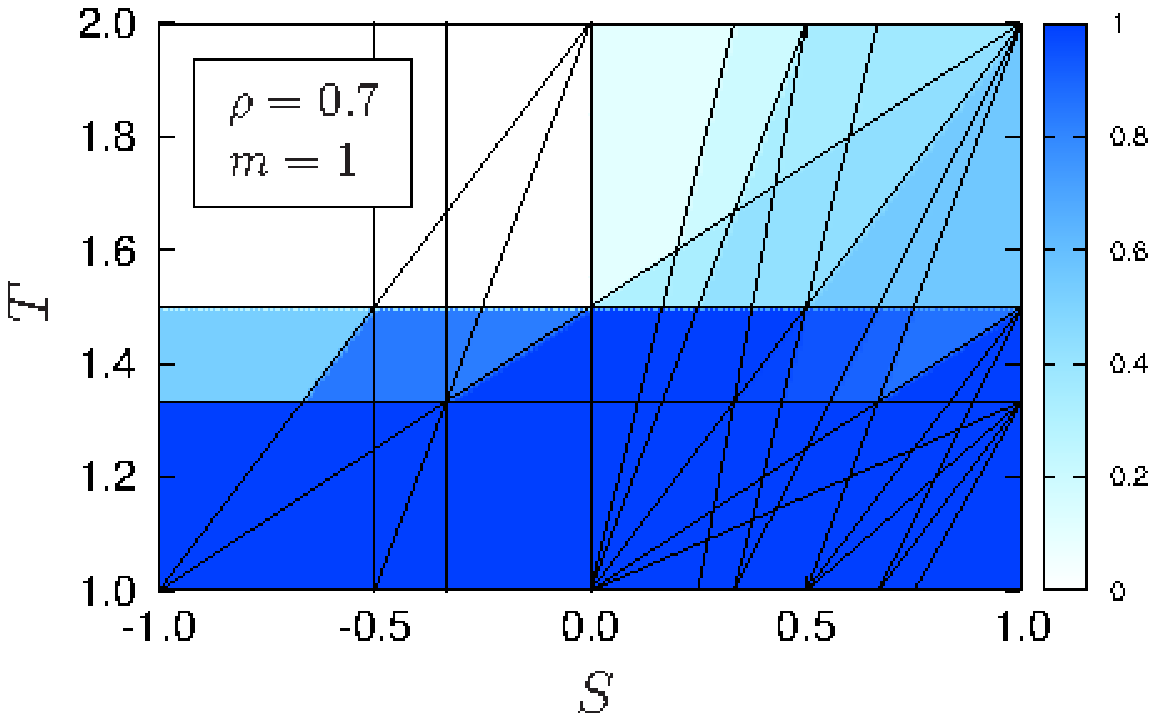}
\caption{Two dimensional cross section of the phase diagram when mobility  is
considered after the combats (CDO dynamics). Notice that although for $m=1$ the density of cooperators is
monotonic, for $m=0.1$ it is not, and intermediate densities sustain more cooperation.}
\label{fig.phase_diagram_CDO}
\end{figure*}

\begin{figure}[htb]
\includegraphics[width=8cm]{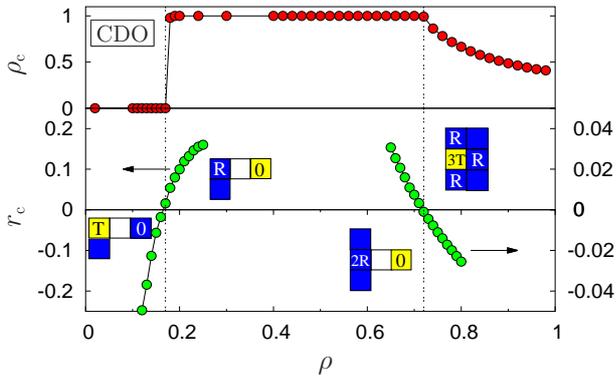}
\caption{Top: Average fraction of cooperating individuals for CDO dynamics,
$m=1$, $T=1.4$ and $S=0$. For intermediate densities, cooperators dominate.
At low densities ($\rho \lesssim 0.18$) there is a discontinuous transition to a phase in
which all agents become defectors while at high densities, $\rho\gtrsim 0.73$, the all
C state continuously turns into a mixed strategy phase. Bottom: order parameters indicating
the leading microscopic processes originating the intermediate all C phase. We also
indicate the more relevant microscopic process in each region. Yellow/light grey are defectors,
blue/dark grey are cooperators, and white boxes are empty sites to which one of the agents may
jump. The letters indicate the payoff accumulated by the combating agents. Notice that the
rightmost process is not diffusive. Below $\rho \simeq 0.18$, since it
has a C neighbor, the defector has payoff $T$ when it collides with the single cooperator whose 
payoff is zero, winning the combat. On the other hand, above $\rho\simeq 0.18$, the cooperator with payoff 
$R$, because of its C neighbor, outperforms the defector whose payoff is
zero. Being the dominant process, this leads to a fast increase in the population of cooperators. Up to
$\rho\simeq 0.73$, cooperators still have an advantage when interacting with defectors after the jump,
since they had accumulated enough payoff from their previous interaction. However, above this value of $\rho$, 
a non diffusive process, reminiscent of the original, full density PD game, allows defectors to invade
previously cooperative regions, thus decreasing $\rhoc$.}
\label{fig.mecanismo_CDO}
\end{figure}

The phase diagram for the CDO (combat-diffusion-offspring) case is slightly different from the one for empty 
sites or COD dynamics. This is due to the fact that in them, the configurations given by $K
_0^{0e_0}$ (D without cooperating neighbors) never compete with any $K_1^{n_1e_1}$ (C with cooperating neighbors) configuration. In the CDO, however, since the
diffusion step occurs between the contest and the generation of offspring, the confrontations
 described above are possible due to the change in configuration between the two steps.
Therefore, by taking these new possibilities into account, a few more lines should 
contribute to the phase diagram, namely
\begin{equation}
g_{(n_0 = 0,n_1,e_0, e_1)}=\frac{(4-e_0)P - n_1 R}{4-(n_1+e_1)},
\label{eq.diag_CDO}
\end{equation}
which are valid for $n_0=0$ and $n_1+e_1 \neq 4$. If the latter inequality is not 
satisfied, the functions will give rise to conditions on $R$ and $P$.  These functions 
mark transitions at $S=g_{(n_0 = 0,n_1,e_0 = 0, e_1)}$, for the different possible values
of $n_1$, $e_1$ and $e_0$.
For $R=1$ and $P=0$, this means that there will be new transition points at the 
following values of $S$: $-3$, $-2$, $-1$, $-1/2$, $-1/3$ and $0$, independent of $e_0$. 
No new transition appears for $S>0$.  It is interesting to notice that due to the new
transition line at $S=0$, this is the only case (besides the mean field) in which the
regions around the weak PD case ($S=0$) differ. These lines are shown in Fig.~\ref{fig.phase_diagram_CDO}
for several values of $\rho$ and $m$. Most of the regions are fully dominated by cooperators, although
a few regions (mainly at small $S$, large $T$) are cooperator free.  As shown in Fig.~\ref{fig.phase_diagram_CDO},
right column, the high mobility case ($m=1$) is more favorable to cooperation at low densities.
On the other hand, for low mobilities (left column), at intermediate densities the fraction of
cooperators is larger. Differently from the COD dynamics, here the agents move after the combat
and thus carry part of their previous history along. This memory of their recent combat is important
to understand the mechanism responsible for the enhancement of cooperation. As an illustration,

 consider the case studied in Ref.~\cite{VaSiAr07}  in which a C dominated phase occurs for $0.18\lesssim \rho \lesssim 0.73$ for $T=1.4$, $S=0$ and $m=1$ (top row of Fig.~\ref{fig.mecanismo_CDO}).
For the low densities close to $\rho \approx 0.18$ the mechanism does not rely on the existence of a spanning
cluster of agents. Indeed, the largest occurrence is of  single and two agent clusters.
The transition rate $\rc$ with which defectors become cooperators depends on the local neighborhood of the two agents at the
moment in which their payoffs were collected (combat) and has two main contributions:
$\rc=\text{Prob}(K_1^{1e_1},K_0^{0e_0})-\text{Prob}(K_1^{0e_1},K_0^{1e_0})$, where the probabilities are the 
number of the given encounter divided by the total number of active sites averaged over time in  
the beginning of the simulation ($ t\leq 100$). That is, the number of cooperators increases
when a C with one C neighbor ($K_1^{1e_1}$) meets, after the jump, a D without C neighbors ($K_0^{0e_0}$)
and decreases when a D also with one C neighbor ($K_0^{1e_0}$) meets, after the jump, 
a C without other C neighbors ($K_1^{0e_1}$). In the bottom row, left side, of Fig.~\ref{fig.mecanismo_CDO}
we depict these processes, each one in the region in which it is dominant, along with the curve showing
that $\rc$ changes sign around the point at which the density of cooperators explodes. What changes from
one case to the other as $\rho$ increases is that clusters with an increasing number of agents become more common
and give support to the stability of the CC pair. 

This transition at low densities does not occur in the COD dynamics and is the mechanism that allows a C to leave a C cluster and continue cooperating in the CDO dynamics. 

The region in which the fraction of cooperators starts to decrease 
($\rho \gtrsim 0.73$) is above the percolation threshold, the system is denser and it is much more likely 
that an individual has several neighbors. Consequently, the microscopic process responsible for this decrease 
is different from the previous one. In effect, the observed decrease in the density of cooperators as 
the lattice approaches full occupancy is rather general, non diffusive and reminiscent of the $m=0$ behavior
for increasing densities~\cite{VaAr01}. Indeed, as the
number of empty sites decreases, irrespective of the mobility, the number of active sites increases,
signaling that there is less defect-induced pinning in the system and cooperators become
more vulnerable to defectors. The mobility $m$, also being a depinning mechanism, plays a role by 
shifting the transition to smaller values of $\rho$.
It should be noted that the main microscopic processes that drive a given transition are dependent on 
the values of $T$ and $S$, besides 
the system occupancy, because a change in $\rho$ modifies the expected number of neighbors that each individual has. 
In the particular case of Fig.~\ref{fig.mecanismo_CDO}, we may also 
point that, differently from the low density transition, there are many microscopic processes that
are relevant for increase of defectors. As an example, consider those associated with neighborhoods $K_1^{2e_1}$
and $K_0^{0e_0}$ ($D\to C$) or $K_1^{1e_1}$ and $K_0^{3e_0}$ ($C \to D$). We show in the right part of
Fig.~\ref{fig.mecanismo_CDO}, the order parameter that can be built from these processes,
$\rc=\text{Prob}(K_1^{2e_1},K_0^{0e_0}) - \text{Prob}(K_1^{1e_1},K_0^{3e_0})$, and how it changes sign
at the transition. There are, however, other possible combinations of microscopic processes 
giving rise to order parameters changing sign in this region (although not necessarily at the same 
precise value).

For the COD dynamics, it is also not easy to single out a few processes that are responsible for changing
the amount of cooperators. 

\begin{figure}[htb]
\includegraphics[width=8cm]{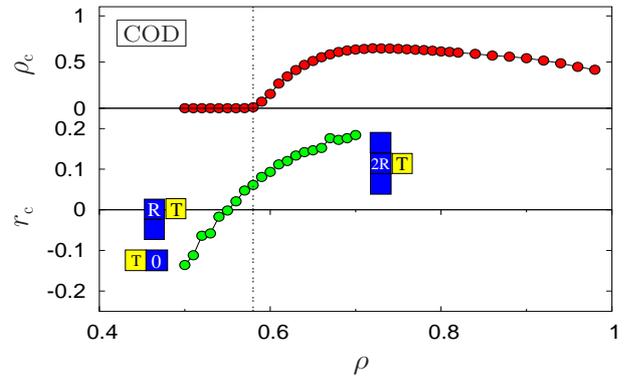}
\caption{Top: Average fraction of cooperating individuals for the COD dynamics,
$m=0.01$, $T=1.4$ and $S=0$. The system presents a transition from a cooperator-free phase to a coexistence phase near 
the percolation threshold $\rho\simeq 0.59$ for the square lattice. Below this density, the evaporation of C clusters
is not counterbalanced by any mechanism and diffusion is detrimental to the population. 
 Bottom: Order parameter associated with the leading microscopic processes that localize the transition
 to the coexistence state. Notice that the processes depicted here are only a few of those existing at
 this density (also explaining why the curve does not cross at the right value).}
\label{fig.mecanismo_COD}
\end{figure}

As an example, we consider the case $T=1.4$, $S=0$ and very low mobility ($m=0.01$)
studied in Ref.~\cite{VaSiAr07}. The larger the mobility, the larger is the density
capable of supporting cooperation. For $m\to 0$, this minimum density seems to 
approach the random site percolation threshold. Above the transition, in the
phase presenting both cooperators and defectors, there is a great chance that 
an individual has $2$ or $3$ neighbors. The main microscopic mechanisms responsible for the 
transition are three encounters: $K_1^{2e_1}$ and $K_0^{1e_0}$ ($D \to C$); $K_1^{1e_1}$ and $K_0^{2e_0}$ ($C \to D$); 
$K_1^{0e_1}$ and $K_0^{1e_0}$ ($C \to D$). The latter corresponds to a C player leaving a C cluster and meeting a D. 
 In Fig.~\ref{fig.mecanismo_COD}, we show the order parameter $\rc=\text{Prob}(K_1^{2e_1},K_0^{1e_0}) - \text{Prob}(K_1^{1e_1},K_0^{2e_0}) - 
\text{Prob}(K_1^{0e_1},K_0^{1e_0})$, 
where the probabilities are calculated as above in the CDO case.

 It takes the value $0$ at $\rho \simeq 0.56$, 
slightly below the transition density. This small difference is due to the fact that many other, less
frequent processes have not been included in the order parameter. 
It should be noted that as the mobility 
probability $m$ increases, the evaporation of C clusters becomes greater and the transition is driven to higher 
values of $\rho$. At these higher concentrations, many other microscopic processes become important to determine the
 value of $\rho$ at which the transition occurs. 

\section{Conclusions}

We presented a systematic study of the Prisoner's Dilemma and Snowdrift
games, with and without random mobility, when the evolution follows a non
stochastic imitation rule in which all agents, in parallel, choose to
follow their more successful neighbor. Few attempts have been made
in the literature~\cite{Hauert01,ScBeMu02,SaPaLe06,ToLuGi06,RoCuSa09a,RoCuSa09b} to 
present a comprehensive account of the possible behaviors of such
evolutionary games. Due to the large number of parameters and possible 
dynamical rules in such models, comparisons among them are difficult
and the universality of the results difficult to access. The phase diagrams and the
corresponding transition lines are obtained by enumerating all
possible local configurations, while the fraction of cooperators in each 
phase is measured in Monte Carlo simulations. Some of the regions 
appearing in these phase diagrams may also be obtained~\cite{Hauert01} 
from the analysis of the fundamental clusters growth conditions.
However, when disorder is present (for example, as dilution), there are
finite size sample to sample fluctuations that depend on the random initial conditions
and the disorder realization. In this case, the cooperative fate of the 
population must be obtained through an average over the disorder and
initial states. Although here we only considered the stationary, asymptotic 
properties of the model, it is interesting to 
notice that the different regions of the phase
diagram may have their dynamic properties characterized by a
Lyapunov exponent~\cite{Hauert01}, the active region where Cs and Ds
coexist being mostly chaotic (positive Lyapunov exponents).

Mobility, in its random flavor considered here, is a stochastic element
that not only prevents the system from being trapped in a frozen state but
may also strongly enhance the amount of cooperation in the system, despite
its tendency to evaporate and disaggregate clusters. Besides exploring, in
the universe of parameters of a particular version of these games, the
conditions under which cooperation may be amplified by spatial and
mobility factors, we also discussed the microscopic mechanisms responsible
for such behavior. Whether a backbone of supporting agents 
is essential, actually depends on the dynamics and the
parameters involved. In fact, the parallel noiseless dynamics considered
here tends to mask the role of the percolating cluster~\cite{WaSzPe12}.
 An intriguing feature is that
instead of smoothing out transitions, when compared with the immobile system,
a few extra transitions are indeed driven by mobility. Interestingly, the
CDO case presents a new transition separating the semi-planes 
$S<0$ (PD) and $S>0$ (SD). The other situation in which this transition 
appears is within mean field. Thus, it differs from the other cases considered
here in which the so called weak version of the PD game presents the same
behavior for both $S=0^+$ and $S=0^-$.

Assuming that the timescale for collecting the payoffs (the combat phase) and
the interval between generating offspring are of the same order, the order
in which these steps are taken becomes relevant. In particular,
the growth mechanism of cooperating clusters differs whether the diffusive step
is performed after or before the reproduction step. In the former (COD), cooperators
have no memory of their previous encounters, carry no payoff while diffusing and
may be easily converted to defectors once they move away from the protective
zone of cooperative clusters. However, diffusing defectors may be assimilated by
these clusters and this passive mechanism may increase cooperation when mobility 
is not extremely high or density is not too low. 
In the latter case (CDO), on the other hand, the agents
have memory of their previous location and may carry a large payoff, thus enabling cooperators to
actively invade regions away from the percolating cluster. The amount of
cooperation in this case is much larger than in the previous one.

In diffusive games, there are two competing parameters: density and diffusivity.
While larger
densities increase the correlation between neighbors, the effect of random 
diffusion depends on the density and sometimes decreases correlation.
It would then be important to compare the results presented here with
those using different updating and diffusion (random) rules, for example,
by allowing multioccupation and position swapping~\cite{SzSzSc08,DrSzSz09}
or different timescales of the selection and fitness collection processes~\cite{GeCrFr13}.

A caveat that must be emphasized concerns the phase diagrams and the fraction of
cooperators measured in the simulations. Although the transition lines are exact,
the densities of cooperators depicted in the previous phase diagrams may slightly 
change when larger system sizes and longer simulation runs are used. Finite size
effects are usually not taken into account once actual populations are
finite both in time and space, but are of interest for a better understanding
of the model.

In summary, we presented a comprehensive study of the PD and SD games under
a deterministic synchronous updating rule in the presence of quenched and
annealed defects. The different phases in the $TS$ plane and the effects of 
dilution and mobility were discussed along with the corresponding microscopic
mechanisms. Under a wide range of conditions, we have shown that mobility,
even if random, may be responsible for a dramatic increase in the population
of cooperators.

\begin{acknowledgments}
Research partially supported by the Brazilian agencies CNPq,
grant PROSUL-490440/2007, CAPES and Fapergs. 
We thank Hugo Fort and Ana T.C. Silva 
for collaborating in subjects related to this project.
 JJA thanks the INCT-Sistemas Complexos (CNPq) for
support and the LPTHE (UPMC, Paris) where part of this paper was written
during his stay. 
\end{acknowledgments}

\bibliographystyle{apsrev4-1}

\begin{thebibliography}{10}%
\makeatletter
\providecommand \@ifxundefined [1]{%
 \ifx #1\undefined \expandafter \@firstoftwo
 \else \expandafter \@secondoftwo
\fi
}%
\providecommand \@ifnum [1]{%
 \ifnum #1\expandafter \@firstoftwo
 \else \expandafter \@secondoftwo
\fi
}%
\providecommand \enquote [1]{``#1''}%
\providecommand \bibnamefont  [1]{#1}%
\providecommand \bibfnamefont [1]{#1}%
\providecommand \citenamefont [1]{#1}%
\providecommand\href[0]{\@sanitize\@href}%
\providecommand\@href[1]{\endgroup\@@startlink{#1}\endgroup\@@href}%
\providecommand\@@href[1]{#1\@@endlink}%
\providecommand \@sanitize [0]{\begingroup\catcode`\&12\catcode`\#12\relax}%
\@ifxundefined \pdfoutput {\@firstoftwo}{%
 \@ifnum{\z@=\pdfoutput}{\@firstoftwo}{\@secondoftwo}%
}{%
 \providecommand\@@startlink[1]{\leavevmode}%
 \providecommand\@@endlink[0]{}%
}{%
 \providecommand\@@startlink[1]{%
  \leavevmode
  \pdfstartlink
   attr{/Border[0 0 1 ]/H/I/C[0 1 1]}%
   user{/Subtype/Link/A<</Type/Action/S/URI/URI(#1)>>}%
  \relax
 }%
 \providecommand\@@endlink[0]{\pdfendlink}%
}%
\providecommand \url  [0]{\begingroup\@sanitize \@url }%
\providecommand \@url [1]{\endgroup\@href {#1}{\urlprefix}}%
\providecommand \urlprefix [0]{URL }%
\providecommand \Eprint[0]{\href }%
\@ifxundefined \urlstyle {%
  \providecommand \doi [1]{doi:\discretionary{}{}{}#1}%
}{%
  \providecommand \doi [0]{doi:\discretionary{}{}{}\begingroup
  \urlstyle{rm}\Url }%
}%
\providecommand \doibase [0]{http://dx.doi.org/}%
\providecommand \Doi[1]{\href{\doibase#1}}%
\providecommand \bibAnnote [3]{%
  \BibitemShut{#1}%
  \begin{quotation}\noindent
    \textsc{Key:}\ #2\\\textsc{Annotation:}\ #3%
  \end{quotation}%
}%
\providecommand \bibAnnoteFile [2]{%
  \IfFileExists{#2}{\bibAnnote {#1} {#2} {\input{#2}}}{}%
}%
\providecommand \typeout [0]{\immediate \write \m@ne }%
\providecommand \selectlanguage [0]{\@gobble}%
\providecommand \bibinfo [0]{\@secondoftwo}%
\providecommand \bibfield [0]{\@secondoftwo}%
\providecommand \translation [1]{[#1]}%
\providecommand \BibitemOpen[0]{}%
\providecommand \bibitemStop [0]{}%
\providecommand \bibitemNoStop [0]{.\EOS\space}%
\providecommand \EOS [0]{\spacefactor3000\relax}%
\providecommand \BibitemShut [1]{\csname bibitem#1\endcsname}%
\bibitem{DoHa05}%
  \BibitemOpen
  \bibfield{author}{%
  \bibinfo {author} {\bibfnamefont{M.}~\bibnamefont{Doebeli}}\ and\ \bibinfo
  {author} {\bibfnamefont{C.}~\bibnamefont{Hauert}},\ }%
  \bibfield{journal}{%
  \bibinfo {journal} {Ecology Letters}\ }%
  \textbf{\bibinfo {volume} {8}},\ \bibinfo {pages} {748} (\bibinfo {year}
  {2005})%
  \bibAnnoteFile{NoStop}{DoHa05}%
\bibitem{Nowak06}%
  \BibitemOpen
  \bibfield{author}{%
  \bibinfo {author} {\bibfnamefont{M.~A.}\ \bibnamefont{Nowak}},\ }%
  \bibfield{journal}{%
  \bibinfo {journal} {Science}\ }%
  \textbf{\bibinfo {volume} {314}},\ \bibinfo {pages} {1560} (\bibinfo {year}
  {2006})%
  \bibAnnoteFile{NoStop}{Nowak06}%
\bibitem{SzFa07}%
  \BibitemOpen
  \bibfield{author}{%
  \bibinfo {author} {\bibfnamefont{G.}~\bibnamefont{Szab\'{o}}}\ and\ \bibinfo
  {author} {\bibfnamefont{G.}~\bibnamefont{F\'{a}th}},\ }%
  \bibfield{journal}{%
  \bibinfo {journal} {Phys. Rep.}\ }%
  \textbf{\bibinfo {volume} {446}},\ \bibinfo {pages} {97} (\bibinfo {year}
  {2007})%
  \bibAnnoteFile{NoStop}{SzFa07}%
\bibitem{RoCuSa09a}%
  \BibitemOpen
  \bibfield{author}{%
  \bibinfo {author} {\bibfnamefont{C.~P.}\ \bibnamefont{Roca}}, \bibinfo
  {author} {\bibfnamefont{J.~A.}\ \bibnamefont{Cuesta}},\ and\ \bibinfo
  {author} {\bibfnamefont{A.}~\bibnamefont{S\'{a}nchez}},\ }%
  \bibfield{journal}{%
  \bibinfo {journal} {Physics of Life Reviews}\ }%
  \textbf{\bibinfo {volume} {6}},\ \bibinfo {pages} {208} (\bibinfo {year}
  {2009})%
  \bibAnnoteFile{NoStop}{RoCuSa09a}%
\bibitem{PeSz10}%
  \BibitemOpen
  \bibfield{author}{%
  \bibinfo {author} {\bibfnamefont{M.}~\bibnamefont{Perc}}\ and\ \bibinfo
  {author} {\bibfnamefont{A.}~\bibnamefont{Szolnoki}},\ }%
  \bibfield{journal}{%
  \bibinfo {journal} {BioSystems}\ }%
  \textbf{\bibinfo {volume} {99}},\ \bibinfo {pages} {109} (\bibinfo {year}
  {2010})%
  \bibAnnoteFile{NoStop}{PeSz10}%
\bibitem{VaAr01}%
  \BibitemOpen
  \bibfield{author}{%
  \bibinfo {author} {\bibfnamefont{M.~H.}\ \bibnamefont{Vainstein}}\ and\
  \bibinfo {author} {\bibfnamefont{J.~J.}\ \bibnamefont{Arenzon}},\ }%
  \bibfield{journal}{%
  \bibinfo {journal} {Phys. Rev. E}\ }%
  \textbf{\bibinfo {volume} {64}},\ \bibinfo {pages} {051905} (\bibinfo {year}
  {2001})%
  \bibAnnoteFile{NoStop}{VaAr01}%
\bibitem{WaSzPe12}%
  \BibitemOpen
  \bibfield{author}{%
  \bibinfo {author} {\bibfnamefont{Z.}~\bibnamefont{Wang}}, \bibinfo {author}
  {\bibfnamefont{A.}~\bibnamefont{Szolnoki}},\ and\ \bibinfo {author}
  {\bibfnamefont{M.}~\bibnamefont{Perc}},\ }%
  \bibfield{journal}{%
  \bibinfo {journal} {Nature Sci. Rep.}\ }%
  \textbf{\bibinfo {volume} {2}},\ \bibinfo {pages} {369} (\bibinfo {year}
  {2012})%
  \bibAnnoteFile{NoStop}{WaSzPe12}%
\bibitem{DuWi91}%
  \BibitemOpen
  \bibfield{author}{%
  \bibinfo {author} {\bibfnamefont{L.~A.}\ \bibnamefont{Dugatkin}}\ and\
  \bibinfo {author} {\bibfnamefont{D.~S.}\ \bibnamefont{Wilson}},\ }%
  \bibfield{journal}{%
  \bibinfo {journal} {Am. Nat.}\ }%
  \textbf{\bibinfo {volume} {138}},\ \bibinfo {pages} {687} (\bibinfo {year}
  {1991})%
  \bibAnnoteFile{NoStop}{DuWi91}%
\bibitem{EnLe93}%
  \BibitemOpen
  \bibfield{author}{%
  \bibinfo {author} {\bibfnamefont{M.}~\bibnamefont{Enquist}}\ and\ \bibinfo
  {author} {\bibfnamefont{O.}~\bibnamefont{Leimar}},\ }%
  \bibfield{journal}{%
  \bibinfo {journal} {Anim. Behav.}\ }%
  \textbf{\bibinfo {volume} {45}},\ \bibinfo {pages} {747} (\bibinfo {year}
  {1993})%
  \bibAnnoteFile{NoStop}{EnLe93}%
\bibitem{FeMi96}%
  \BibitemOpen
  \bibfield{author}{%
  \bibinfo {author} {\bibfnamefont{R.}~\bibnamefont{Ferri\`{e}re}}\ and\
  \bibinfo {author} {\bibfnamefont{R.~E.}\ \bibnamefont{Michod}},\ }%
  \bibfield{journal}{%
  \bibinfo {journal} {Am. Nat.}\ }%
  \textbf{\bibinfo {volume} {147}},\ \bibinfo {pages} {692} (\bibinfo {year}
  {1996})%
  \bibAnnoteFile{NoStop}{FeMi96}%
\bibitem{HaTa05}%
  \BibitemOpen
  \bibfield{author}{%
  \bibinfo {author} {\bibfnamefont{I.~M.}\ \bibnamefont{Hamilton}}\ and\
  \bibinfo {author} {\bibfnamefont{M.}~\bibnamefont{Taborsky}},\ }%
  \bibfield{journal}{%
  \bibinfo {journal} {Proc. R. Soc. B}\ }%
  \textbf{\bibinfo {volume} {272}},\ \bibinfo {pages} {2259} (\bibinfo {year}
  {2005})%
  \bibAnnoteFile{NoStop}{HaTa05}%
\bibitem{LeFeDi05}%
  \BibitemOpen
  \bibfield{author}{%
  \bibinfo {author} {\bibfnamefont{J.-F.}\ \bibnamefont{{Le Galliard}}},
  \bibinfo {author} {\bibfnamefont{F.}~\bibnamefont{Ferri\`{e}re}},\ and\
  \bibinfo {author} {\bibfnamefont{U.}~\bibnamefont{Dieckmann}},\ }%
  \bibfield{journal}{%
  \bibinfo {journal} {Am. Nat.}\ }%
  \textbf{\bibinfo {volume} {165}},\ \bibinfo {pages} {206} (\bibinfo {year}
  {2005})%
  \bibAnnoteFile{NoStop}{LeFeDi05}%
\bibitem{VaSiAr07}%
  \BibitemOpen
  \bibfield{author}{%
  \bibinfo {author} {\bibfnamefont{M.~H.}\ \bibnamefont{Vainstein}}, \bibinfo
  {author} {\bibfnamefont{A.~T.~C.}\ \bibnamefont{Silva}},\ and\ \bibinfo
  {author} {\bibfnamefont{J.~J.}\ \bibnamefont{Arenzon}},\ }%
  \bibfield{journal}{%
  \bibinfo {journal} {J. Theor. Biol.}\ }%
  \textbf{\bibinfo {volume} {244}},\ \bibinfo {pages} {722} (\bibinfo {year}
  {2007})%
  \bibAnnoteFile{NoStop}{VaSiAr07}%
\bibitem{JiZhYi07}%
  \BibitemOpen
  \bibfield{author}{%
  \bibinfo {author} {\bibfnamefont{G.}~\bibnamefont{Jian-Yue}}, \bibinfo
  {author} {\bibfnamefont{W.}~\bibnamefont{Zhi-Xi}},\ and\ \bibinfo {author}
  {\bibfnamefont{W.}~\bibnamefont{Ying-Hai}},\ }%
  \bibfield{journal}{%
  \bibinfo {journal} {Chin. Phys.}\ }%
  \textbf{\bibinfo {volume} {16}},\ \bibinfo {pages} {3566} (\bibinfo {year}
  {2007})%
  \bibAnnoteFile{NoStop}{JiZhYi07}%
\bibitem{SzSzSc08}%
  \BibitemOpen
  \bibfield{author}{%
  \bibinfo {author} {\bibfnamefont{S.}~\bibnamefont{Sz\'{a}mad\'{o}}}, \bibinfo
  {author} {\bibfnamefont{F.}~\bibnamefont{Szalai}},\ and\ \bibinfo {author}
  {\bibfnamefont{I.}~\bibnamefont{Scheuring}},\ }%
  \bibfield{journal}{%
  \bibinfo {journal} {J. Theor. Biol.}\ }%
  \textbf{\bibinfo {volume} {253}},\ \bibinfo {pages} {221} (\bibinfo {year}
  {2008})%
  \bibAnnoteFile{NoStop}{SzSzSc08}%
\bibitem{DrSzSz09}%
  \BibitemOpen
  \bibfield{author}{%
  \bibinfo {author} {\bibfnamefont{M.}~\bibnamefont{Droz}}, \bibinfo {author}
  {\bibfnamefont{J.}~\bibnamefont{Szwabiński}},\ and\ \bibinfo {author}
  {\bibfnamefont{G.}~\bibnamefont{Szab\'{o}}},\ }%
  \bibfield{journal}{%
  \bibinfo {journal} {Eur. Phys. J. B}\ }%
  \textbf{\bibinfo {volume} {71}},\ \bibinfo {pages} {579} (\bibinfo {year}
  {2009})%
  \bibAnnoteFile{NoStop}{DrSzSz09}%
\bibitem{SiFoVaAr09}%
  \BibitemOpen
  \bibfield{author}{%
  \bibinfo {author} {\bibfnamefont{E.~A.}\ \bibnamefont{Sicardi}}, \bibinfo
  {author} {\bibfnamefont{H.}~\bibnamefont{Fort}}, \bibinfo {author}
  {\bibfnamefont{M.~H.}\ \bibnamefont{Vainstein}},\ and\ \bibinfo {author}
  {\bibfnamefont{J.~J.}\ \bibnamefont{Arenzon}},\ }%
  \bibfield{journal}{%
  \bibinfo {journal} {J. Theor. Biol.}\ }%
  \textbf{\bibinfo {volume} {256}},\ \bibinfo {pages} {240} (\bibinfo {year}
  {2009})%
  \bibAnnoteFile{NoStop}{SiFoVaAr09}%
\bibitem{YaWa11}%
  \BibitemOpen
  \bibfield{author}{%
  \bibinfo {author} {\bibfnamefont{H.}~\bibnamefont{Yang}}\ and\ \bibinfo
  {author} {\bibfnamefont{B.}~\bibnamefont{Wang}},\ }%
  \bibfield{journal}{%
  \bibinfo {journal} {Chin. Sci. Bull.}\ }%
  \textbf{\bibinfo {volume} {56}},\ \bibinfo {pages} {3693} (\bibinfo {year}
  {2011})%
  \bibAnnoteFile{NoStop}{YaWa11}%
\bibitem{SuKi11}%
  \BibitemOpen
  \bibfield{author}{%
  \bibinfo {author} {\bibfnamefont{S.}~\bibnamefont{Suzuki}}\ and\ \bibinfo
  {author} {\bibfnamefont{H.}~\bibnamefont{Kimura}},\ }%
  \bibfield{journal}{%
  \bibinfo {journal} {J. Theor. Biol.}\ }%
  \textbf{\bibinfo {volume} {287}},\ \bibinfo {pages} {42} (\bibinfo {year}
  {2011})%
  \bibAnnoteFile{NoStop}{SuKi11}%
\bibitem{GeCrFr13}%
  \BibitemOpen
  \bibfield{author}{%
  \bibinfo {author} {\bibfnamefont{A.}~\bibnamefont{Gelimson}}, \bibinfo
  {author} {\bibfnamefont{J.}~\bibnamefont{Cremer}},\ and\ \bibinfo {author}
  {\bibfnamefont{E.}~\bibnamefont{Frey}},\ }%
  \bibfield{journal}{%
  \bibinfo {journal} {Phys. Rev. E}\ }%
  \textbf{\bibinfo {volume} {87}},\ \bibinfo {pages} {042711} (\bibinfo {year}
  {2013})%
  \bibAnnoteFile{NoStop}{GeCrFr13}%
\bibitem{ChLiDaZhYa10}%
  \BibitemOpen
  \bibfield{author}{%
  \bibinfo {author} {\bibfnamefont{H.}~\bibnamefont{Cheng}}, \bibinfo {author}
  {\bibfnamefont{H.}~\bibnamefont{Li}}, \bibinfo {author}
  {\bibfnamefont{Q.}~\bibnamefont{Dai}}, \bibinfo {author}
  {\bibfnamefont{Y.}~\bibnamefont{Zhu}},\ and\ \bibinfo {author}
  {\bibfnamefont{J.}~\bibnamefont{Yang}},\ }%
  \bibfield{journal}{%
  \bibinfo {journal} {New J. Phys.}\ }%
  \textbf{\bibinfo {volume} {12}},\ \bibinfo {pages} {123014} (\bibinfo {year}
  {2010})%
  \bibAnnoteFile{NoStop}{ChLiDaZhYa10}%
\bibitem{YaWuWa10}%
  \BibitemOpen
  \bibfield{author}{%
  \bibinfo {author} {\bibfnamefont{H.-X.}\ \bibnamefont{Yang}}, \bibinfo
  {author} {\bibfnamefont{Z.-X.}\ \bibnamefont{Wu}},\ and\ \bibinfo {author}
  {\bibfnamefont{B.-H.}\ \bibnamefont{Wang}},\ }%
  \bibfield{journal}{%
  \bibinfo {journal} {Phys. Rev. E}\ }%
  \textbf{\bibinfo {volume} {81}},\ \bibinfo {pages} {065101(R)} (\bibinfo
  {year} {2010})%
  \bibAnnoteFile{NoStop}{YaWuWa10}%
\bibitem{ChDaLiZhZhYa11}%
  \BibitemOpen
  \bibfield{author}{%
  \bibinfo {author} {\bibfnamefont{H.}~\bibnamefont{Cheng}}, \bibinfo {author}
  {\bibfnamefont{Q.}~\bibnamefont{Dai}}, \bibinfo {author}
  {\bibfnamefont{H.}~\bibnamefont{Li}}, \bibinfo {author}
  {\bibfnamefont{Y.}~\bibnamefont{Zhu}}, \bibinfo {author}
  {\bibfnamefont{M.}~\bibnamefont{Zhang}},\ and\ \bibinfo {author}
  {\bibfnamefont{J.}~\bibnamefont{Yang}},\ }%
  \bibfield{journal}{%
  \bibinfo {journal} {New J. Phys.}\ }%
  \textbf{\bibinfo {volume} {13}},\ \bibinfo {pages} {043032} (\bibinfo {year}
  {2011})%
  \bibAnnoteFile{NoStop}{ChDaLiZhZhYa11}%
\bibitem{LiYaSh11}%
  \BibitemOpen
  \bibfield{author}{%
  \bibinfo {author} {\bibfnamefont{H.}~\bibnamefont{Lin}}, \bibinfo {author}
  {\bibfnamefont{D.-P.}\ \bibnamefont{Yang}},\ and\ \bibinfo {author}
  {\bibfnamefont{J.}~\bibnamefont{Shuai}},\ }%
  \bibfield{journal}{%
  \bibinfo {journal} {Chaos, Solitons \& Fractals}\ }%
  \textbf{\bibinfo {volume} {44}},\ \bibinfo {pages} {153} (\bibinfo {year}
  {2011})%
  \bibAnnoteFile{NoStop}{LiYaSh11}%
\bibitem{LiYaWuWa11}%
  \BibitemOpen
  \bibfield{author}{%
  \bibinfo {author} {\bibfnamefont{Y.-T.}\ \bibnamefont{Lin}}, \bibinfo
  {author} {\bibfnamefont{H.-X.}\ \bibnamefont{Yang}}, \bibinfo {author}
  {\bibfnamefont{Z.-X.}\ \bibnamefont{Wu}},\ and\ \bibinfo {author}
  {\bibfnamefont{B.-H.}\ \bibnamefont{Wang}},\ }%
  \bibfield{journal}{%
  \bibinfo {journal} {Physica A}\ }%
  \textbf{\bibinfo {volume} {390}},\ \bibinfo {pages} {77} (\bibinfo {year}
  {2011})%
  \bibAnnoteFile{NoStop}{LiYaWuWa11}%
\bibitem{HeYu09}%
  \BibitemOpen
  \bibfield{author}{%
  \bibinfo {author} {\bibfnamefont{D.}~\bibnamefont{Helbing}}\ and\ \bibinfo
  {author} {\bibfnamefont{W.}~\bibnamefont{Yu}},\ }%
  \bibfield{journal}{%
  \bibinfo {journal} {Proc. Nat. Acad. Sci.}\ }%
  \textbf{\bibinfo {volume} {106}},\ \bibinfo {pages} {3680} (\bibinfo {year}
  {2009})%
  \bibAnnoteFile{NoStop}{HeYu09}%
\bibitem{Helbing09}%
  \BibitemOpen
  \bibfield{author}{%
  \bibinfo {author} {\bibfnamefont{D.}~\bibnamefont{Helbing}},\ }%
  \bibfield{journal}{%
  \bibinfo {journal} {Eur. Phys. J. B}\ }%
  \textbf{\bibinfo {volume} {67}},\ \bibinfo {pages} {345} (\bibinfo {year}
  {2009})%
  \bibAnnoteFile{NoStop}{Helbing09}%
\bibitem{Yu11}%
  \BibitemOpen
  \bibfield{author}{%
  \bibinfo {author} {\bibfnamefont{W.}~\bibnamefont{Yu}},\ }%
  \bibfield{journal}{%
  \bibinfo {journal} {Phys. Rev. E}\ }%
  \textbf{\bibinfo {volume} {83}},\ \bibinfo {pages} {026105} (\bibinfo {year}
  {2011})%
  \bibAnnoteFile{NoStop}{Yu11}%
\bibitem{Aktipis04}%
  \BibitemOpen
  \bibfield{author}{%
  \bibinfo {author} {\bibfnamefont{C.~A.}\ \bibnamefont{Aktipis}},\ }%
  \bibfield{journal}{%
  \bibinfo {journal} {J. Theor. Biol.}\ }%
  \textbf{\bibinfo {volume} {231}},\ \bibinfo {pages} {249} (\bibinfo {year}
  {2004})%
  \bibAnnoteFile{NoStop}{Aktipis04}%
\bibitem{JiWaLaWa10}%
  \BibitemOpen
  \bibfield{author}{%
  \bibinfo {author} {\bibfnamefont{L.-L.}\ \bibnamefont{Jiang}}, \bibinfo
  {author} {\bibfnamefont{W.-X.}\ \bibnamefont{Wang}}, \bibinfo {author}
  {\bibfnamefont{Y.-C.}\ \bibnamefont{Lai}},\ and\ \bibinfo {author}
  {\bibfnamefont{B.-H.}\ \bibnamefont{Wang}},\ }%
  \bibfield{journal}{%
  \bibinfo {journal} {Phys. Rev. E}\ }%
  \textbf{\bibinfo {volume} {81}},\ \bibinfo {pages} {036108} (\bibinfo {year}
  {2010})%
  \bibAnnoteFile{NoStop}{JiWaLaWa10}%
\bibitem{Aktipis11}%
  \BibitemOpen
  \bibfield{author}{%
  \bibinfo {author} {\bibfnamefont{C.~A.}\ \bibnamefont{Aktipis}},\ }%
  \bibfield{journal}{%
  \bibinfo {journal} {Evol. Hum. Behav.}\ }%
  \textbf{\bibinfo {volume} {32}},\ \bibinfo {pages} {263} (\bibinfo {year}
  {2011})%
  \bibAnnoteFile{NoStop}{Aktipis11}%
\bibitem{ChGaCaXu11a}%
  \BibitemOpen
  \bibfield{author}{%
  \bibinfo {author} {\bibfnamefont{Z.}~\bibnamefont{Chen}}, \bibinfo {author}
  {\bibfnamefont{J.}~\bibnamefont{Gao}}, \bibinfo {author}
  {\bibfnamefont{Y.}~\bibnamefont{Cai}},\ and\ \bibinfo {author}
  {\bibfnamefont{X.}~\bibnamefont{Xu}},\ }%
  \bibfield{journal}{%
  \bibinfo {journal} {Physica A}\ }%
  \textbf{\bibinfo {volume} {390}},\ \bibinfo {pages} {1615} (\bibinfo {year}
  {2011})%
  \bibAnnoteFile{NoStop}{ChGaCaXu11a}%
\bibitem{ZhWaDuCa11}%
  \BibitemOpen
  \bibfield{author}{%
  \bibinfo {author} {\bibfnamefont{J.}~\bibnamefont{Zhang}}, \bibinfo {author}
  {\bibfnamefont{W.-Y.}\ \bibnamefont{Wang}}, \bibinfo {author}
  {\bibfnamefont{W.-B.}\ \bibnamefont{Du}},\ and\ \bibinfo {author}
  {\bibfnamefont{X.-B.}\ \bibnamefont{Cao}},\ }%
  \bibfield{journal}{%
  \bibinfo {journal} {Physica A}\ }%
  \textbf{\bibinfo {volume} {390}},\ \bibinfo {pages} {2251} (\bibinfo {year}
  {2011})%
  \bibAnnoteFile{NoStop}{ZhWaDuCa11}%
\bibitem{ZhZhWePeXiWa12}%
  \BibitemOpen
  \bibfield{author}{%
  \bibinfo {author} {\bibfnamefont{C.}~\bibnamefont{Zhang}}, \bibinfo {author}
  {\bibfnamefont{J.}~\bibnamefont{Zhang}}, \bibinfo {author}
  {\bibfnamefont{F.~J.}\ \bibnamefont{Weissing}}, \bibinfo {author}
  {\bibfnamefont{M.}~\bibnamefont{Perc}}, \bibinfo {author}
  {\bibfnamefont{G.}~\bibnamefont{Xie}},\ and\ \bibinfo {author}
  {\bibfnamefont{L.}~\bibnamefont{Wang}},\ }%
  \bibfield{journal}{%
  \bibinfo {journal} {PloS one}\ }%
  \textbf{\bibinfo {volume} {7}},\ \bibinfo {pages} {e35183} (\bibinfo {year}
  {2012})%
  \bibAnnoteFile{NoStop}{ZhZhWePeXiWa12}%
\bibitem{MeBuFoFrGoLaMo09}%
  \BibitemOpen
  \bibfield{author}{%
  \bibinfo {author} {\bibfnamefont{S.}~\bibnamefont{Meloni}}, \bibinfo {author}
  {\bibfnamefont{A.}~\bibnamefont{Buscarino}}, \bibinfo {author}
  {\bibfnamefont{L.}~\bibnamefont{Fortuna}}, \bibinfo {author}
  {\bibfnamefont{M.}~\bibnamefont{Frasca}}, \bibinfo {author}
  {\bibfnamefont{J.}~\bibnamefont{G\'{o}mez-Garde\~{n}es}}, \bibinfo {author}
  {\bibfnamefont{V.}~\bibnamefont{Latora}},\ and\ \bibinfo {author}
  {\bibfnamefont{Y.}~\bibnamefont{Moreno}},\ }%
  \bibfield{journal}{%
  \bibinfo {journal} {Phys. Rev. E}\ }%
  \textbf{\bibinfo {volume} {79}},\ \bibinfo {pages} {067101} (\bibinfo {year}
  {2009})%
  \bibAnnoteFile{NoStop}{MeBuFoFrGoLaMo09}%
\bibitem{Koella00}%
  \BibitemOpen
  \bibfield{author}{%
  \bibinfo {author} {\bibfnamefont{J.~C.}\ \bibnamefont{Koella}},\ }%
  \bibfield{journal}{%
  \bibinfo {journal} {Proc. R. Soc. B}\ }%
  \textbf{\bibinfo {volume} {267}},\ \bibinfo {pages} {1979} (\bibinfo {year}
  {2000})%
  \bibAnnoteFile{NoStop}{Koella00}%
\bibitem{Rapoport66}%
  \BibitemOpen
  \bibfield{author}{%
  \bibinfo {author} {\bibfnamefont{A.}~\bibnamefont{Rapoport}},\ }%
  \emph{\bibinfo {title} {Two-Person Game Theory}}\ (\bibinfo {publisher} {U.
  of Michigan},\ \bibinfo {address} {Ann Harbor},\ \bibinfo {year} {1966})%
  \bibAnnoteFile{NoStop}{Rapoport66}%
\bibitem{NoMa92}%
  \BibitemOpen
  \bibfield{author}{%
  \bibinfo {author} {\bibfnamefont{M.~A.}\ \bibnamefont{Nowak}}\ and\ \bibinfo
  {author} {\bibfnamefont{R.~M.}\ \bibnamefont{May}},\ }%
  \bibfield{journal}{%
  \bibinfo {journal} {Nature}\ }%
  \textbf{\bibinfo {volume} {359}},\ \bibinfo {pages} {826} (\bibinfo {year}
  {1992})%
  \bibAnnoteFile{NoStop}{NoMa92}%
\bibitem{RoCuSa09b}%
  \BibitemOpen
  \bibfield{author}{%
  \bibinfo {author} {\bibfnamefont{C.~P.}\ \bibnamefont{Roca}}, \bibinfo
  {author} {\bibfnamefont{J.~A.}\ \bibnamefont{Cuesta}},\ and\ \bibinfo
  {author} {\bibfnamefont{A.}~\bibnamefont{S\'{a}nchez}},\ }%
  \bibfield{journal}{%
  \bibinfo {journal} {Phys. Rev. E}\ }%
  \textbf{\bibinfo {volume} {80}},\ \bibinfo {pages} {046106} (\bibinfo {year}
  {2009})%
  \bibAnnoteFile{NoStop}{RoCuSa09b}%
\bibitem{NoMa93}%
  \BibitemOpen
  \bibfield{author}{%
  \bibinfo {author} {\bibfnamefont{M.~A.}\ \bibnamefont{Nowak}}\ and\ \bibinfo
  {author} {\bibfnamefont{R.~M.}\ \bibnamefont{May}},\ }%
  \bibfield{journal}{%
  \bibinfo {journal} {Int. J. of Bifurcation and Chaos}\ }%
  \textbf{\bibinfo {volume} {3}},\ \bibinfo {pages} {35} (\bibinfo {year}
  {1993})%
  \bibAnnoteFile{NoStop}{NoMa93}%
\bibitem{ScBeMu02}%
  \BibitemOpen
  \bibfield{author}{%
  \bibinfo {author} {\bibfnamefont{F.}~\bibnamefont{Schweitzer}}, \bibinfo
  {author} {\bibfnamefont{L.}~\bibnamefont{Behera}},\ and\ \bibinfo {author}
  {\bibfnamefont{H.}~\bibnamefont{Muhlenbein}},\ }%
  \bibfield{journal}{%
  \bibinfo {journal} {Adv. Comp. Syst.}\ }%
  \textbf{\bibinfo {volume} {5}},\ \bibinfo {pages} {269} (\bibinfo {year}
  {2002})%
  \bibAnnoteFile{NoStop}{ScBeMu02}%
\bibitem{Hauert02}%
  \BibitemOpen
  \bibfield{author}{%
  \bibinfo {author} {\bibfnamefont{C.}~\bibnamefont{Hauert}},\ }%
  \bibfield{journal}{%
  \bibinfo {journal} {Int. J. Bif. Chaos}\ }%
  \textbf{\bibinfo {volume} {12}},\ \bibinfo {pages} {1531} (\bibinfo {year}
  {2002})%
  \bibAnnoteFile{NoStop}{Hauert02}%
\bibitem{SaPaLe06}%
  \BibitemOpen
  \bibfield{author}{%
  \bibinfo {author} {\bibfnamefont{F.~C.}\ \bibnamefont{Santos}}, \bibinfo
  {author} {\bibfnamefont{J.~M.}\ \bibnamefont{Pacheco}},\ and\ \bibinfo
  {author} {\bibfnamefont{T.}~\bibnamefont{Lenaerts}},\ }%
  \bibfield{journal}{%
  \bibinfo {journal} {Proc. Natl. Acad. Sci.}\ }%
  \textbf{\bibinfo {volume} {103}},\ \bibinfo {pages} {3490} (\bibinfo {year}
  {2006})%
  \bibAnnoteFile{NoStop}{SaPaLe06}%
\bibitem{ToLuGi06}%
  \BibitemOpen
  \bibfield{author}{%
  \bibinfo {author} {\bibfnamefont{M.}~\bibnamefont{Tomassini}}, \bibinfo
  {author} {\bibfnamefont{L.}~\bibnamefont{Luthi}},\ and\ \bibinfo {author}
  {\bibfnamefont{M.}~\bibnamefont{Giacobini}},\ }%
  \bibfield{journal}{%
  \bibinfo {journal} {Phys. Rev. E}\ }%
  \textbf{\bibinfo {volume} {73}},\ \bibinfo {pages} {016132} (\bibinfo {year}
  {2006})%
  \bibAnnoteFile{NoStop}{ToLuGi06}%
\bibitem{LiKeLiHu12}%
  \BibitemOpen
  \bibfield{author}{%
  \bibinfo {author} {\bibfnamefont{P.-P.}\ \bibnamefont{Li}}, \bibinfo {author}
  {\bibfnamefont{J.}~\bibnamefont{Ke}}, \bibinfo {author}
  {\bibfnamefont{Z.}~\bibnamefont{Lin}},\ and\ \bibinfo {author}
  {\bibfnamefont{P.~M.}\ \bibnamefont{Hui}},\ }%
  \bibfield{journal}{%
  \bibinfo {journal} {Phys. Rev. E}\ }%
  \textbf{\bibinfo {volume} {85}},\ \bibinfo {pages} {021111} (\bibinfo {year}
  {2012})%
  \bibAnnoteFile{NoStop}{LiKeLiHu12}%
\bibitem{Note1}%
  \BibitemOpen
  \bibinfo {note} {There are indeed two more regions for $T<1$, but they are
  not considered here: for $S>0$ (Harmony game), $\rho _{\scriptscriptstyle
  \protect \rm c}=1$, and for $S<0$ (Stag Hunt game), the amount of cooperators
  depends on their initial density: if it is larger (smaller) than $S/(S+T-1)$,
  then $\rho _{\scriptscriptstyle \protect \rm c}=0$ (1).}%
  \bibAnnoteFile{Stop}{Note1}%
\bibitem{Note2}%
  \BibitemOpen
  \bibinfo {note} {In this paper, all densities are relative to the total
  number of agents, not sites.}%
  \bibAnnoteFile{Stop}{Note2}%
\bibitem{Hauert01}%
  \BibitemOpen
  \bibfield{author}{%
  \bibinfo {author} {\bibfnamefont{C.}~\bibnamefont{Hauert}},\ }%
  \bibfield{journal}{%
  \bibinfo {journal} {Proc. R. Soc. Lond. B}\ }%
  \textbf{\bibinfo {volume} {268}},\ \bibinfo {pages} {761} (\bibinfo {year}
  {2001})%
  \bibAnnoteFile{NoStop}{Hauert01}%
\bibitem{HaDo04}%
  \BibitemOpen
  \bibfield{author}{%
  \bibinfo {author} {\bibfnamefont{C.}~\bibnamefont{Hauert}}\ and\ \bibinfo
  {author} {\bibfnamefont{M.}~\bibnamefont{Doebeli}},\ }%
  \bibfield{journal}{%
  \bibinfo {journal} {Nature}\ }%
  \textbf{\bibinfo {volume} {428}},\ \bibinfo {pages} {643} (\bibinfo {year}
  {2004})%
  \bibAnnoteFile{NoStop}{HaDo04}%
\bibitem{SzSzVaHa10}%
  \BibitemOpen
  \bibfield{author}{%
  \bibinfo {author} {\bibfnamefont{G.}~\bibnamefont{Szabó}}, \bibinfo {author}
  {\bibfnamefont{A.}~\bibnamefont{Szolnoki}}, \bibinfo {author}
  {\bibfnamefont{M.}~\bibnamefont{Varga}},\ and\ \bibinfo {author}
  {\bibfnamefont{L.}~\bibnamefont{Hanusovszky}},\ }%
  \bibfield{journal}{%
  \bibinfo {journal} {Phys. Rev. E}\ }%
  \textbf{\bibinfo {volume} {82}},\ \bibinfo {pages} {026110} (\bibinfo {year}
  {2010})%
  \bibAnnoteFile{NoStop}{SzSzVaHa10}%
\bibitem{LiLiTiSh10}%
  \BibitemOpen
  \bibfield{author}{%
  \bibinfo {author} {\bibfnamefont{M.}~\bibnamefont{Lin}}, \bibinfo {author}
  {\bibfnamefont{N.}~\bibnamefont{Li}}, \bibinfo {author}
  {\bibfnamefont{L.}~\bibnamefont{Tian}},\ and\ \bibinfo {author}
  {\bibfnamefont{D.-N.}\ \bibnamefont{Shi}},\ }%
  \bibfield{journal}{%
  \bibinfo {journal} {Physica A}\ }%
  \textbf{\bibinfo {volume} {389}},\ \bibinfo {pages} {1753} (\bibinfo {year}
  {2010})%
  \bibAnnoteFile{NoStop}{LiLiTiSh10}%
\end{thebibliography}

%

\end{document}